%% file: satellites.tex
\documentclass[twocolumn,apj]{openjournal}
\let\tablenum\relax
\usepackage{siunitx}
\usepackage{amsmath}
\usepackage{longtable}
\usepackage{graphicx}
\usepackage{booktabs}
\usepackage{appendix}
\usepackage{url}
\usepackage{hyperref}

\graphicspath{{FIGURES/}}
\begin{document}

\title{Detection of Thermal Emission at Millimeter Wavelengths from Low-Earth Orbit Satellites }

\input{authors_oja}
\thanks{$^{\star}$E-mail:\email{amfoster@princeton.edu}}
\shortauthors{Foster et al.}
\shorttitle{mm-wave Thermal Emission from LEO Satellites}

\begin{abstract}
The detection of artificial satellite thermal emission at millimeter wavelengths is presented using data from the 3rd-Generation receiver on the South Pole Telescope (SPT-3G).
This represents the first reported detection of thermal emission from artificial satellites at millimeter wavelengths.
Satellite thermal emission is shown to be detectable at high signal-to-noise ratios on timescales as short as a few tens of milliseconds.
An algorithm for downloading orbital information and tracking known satellites given observer constraints and time-ordered observatory pointing is described.
Consequences for cosmological surveys and short-duration transient searches are discussed, revealing that the integrated thermal emission from all large satellites does not contribute significantly to the SPT-3G survey intensity map.
Measured satellite positions are found to be discrepant from their two-line element (TLE) derived ephemerides up to several arcminutes which may present a difficulty in cross-checking or masking satellites from short-duration transient searches.

\keywords{satellites, thermal emission, millimeter, time domain astronomy, cosmic microwave background radiation}
\end{abstract}

\section{Introduction}
\label{sec:intro}
As cosmic microwave background (CMB) radiation survey experiments become increasingly sensitive, and with the great community interest in multi-wavelength astronomy, these experiments have begun to transcend their role as observers of a static sky and are now pushing the boundaries of time-domain astrophysics \citep{whitehorn16,naess21,guns21,chichura22,li23,tandoi24,biermann24}.
Because of the challenges in achieving high angular resolution for large-field surveys in the millimeter-wavelength (mm-wave) regime, there are a limited number of such surveys that prioritize time-domain science.
While the South Pole Telescope (SPT) is primarily designed to measure anisotropies in the CMB, due to the 10 meter primary mirror, large field of view, sub-Jansky single-visit sensitivity, and short re-observation cadence, it has become an active contributor to survey-based mm-wave time-domain science \citep{whitehorn16,guns21,hood23,tandoi24}.

Some of the most energetic and interesting time-variable phenomena in the universe occur on timescales of hours or shorter.
These include, but are not limited to, gamma-ray burst reverse shocks, neutron star mergers, X-ray binaries, active galactic nuclei jets, stellar flares, and fast-radio bursts \citep{laskar13,laskar19,macgregor20,CHIME21}.
As many of these events are extremely rare, fast, or faint in the mm-wave, it is imperative to understand the false-positive noise population.

In this context, satellites present a significant challenge as they may potentially appear as false-positive short-duration transients in the time-resolved survey data. 
In contrast to optical surveys, where contamination from reflections of sunlight or Earthshine dominates, at CMB survey wavelengths the thermal emission from the satellites themselves may cause short, high signal-to-noise foreground signals.
Due to the passive nature of the thermal emission, this paper makes no distinction between the various types of resident space objects, referring to active payloads, defunct satellites, and debris such as rocket bodies all simply as ``satellites".

As transient detection algorithms used in CMB-surveys are advanced, it is likely that they will be sending out public alerts on timescales as short as a few hours.
With an increasing number of satellites in orbit, particularly those in low Earth orbit (LEO), it is increasingly likely that they will be mistaken for an astrophysical transient if not correctly identified.
In addition to passive thermal emission, it is common for active payloads to transmit at microwave frequencies \citep{yost23}, potentially further complicating foreground noise.

This paper describes a simple model for the thermal mm-emission of LEO satellites and confirms the accuracy of this model with measurements made by the SPT-3G instrument. 
Section \ref{sec:spt3g} contains details of the SPT-3G instrument, including the observing strategy and survey regions, with particular attention to the properties of time-ordered data pertaining to the visibility of fast-moving satellites.
In Section \ref{sec:satellites}, a brief discussion of the plethora of satellites in orbit around Earth is followed by the thermal model expectations for millimeter emission. 
A description of the algorithms for detecting and measuring satellite fluxes is presented in Section \ref{sec:detecting_satellites}.
Several examples of measured satellite emission are presented in Section \ref{sec:results}, covering both passive and intended emission. 
The integrated signal from all satellites in the survey region that are predicted to be above the instantaneous noise level is extrapolated and the effects on map depth are presented in Section \ref{sec:effects_on_cmb_surveys}. 
The paper concludes with a discussion of the results and a note of caution for future short-duration transient searches.

\section{The South Pole Telescope}
\label{sec:spt3g}
The SPT is a 10-meter diameter primary aperture, mm-wave telescope that has been in operation at the Amundsen-Scott South Pole station since 2007. 
The telescope is physically located approximately 1~km from the geographical South Pole, at Latitude 89d 59.38$'$ S, Longitude 44d 12.08$'$ W, Altitude 2818~meters.
The SPT has been used almost exclusively to make deep maps of thousands of square degrees of the southern sky, with the primary science goal of characterizing the primary and secondary CMB anisotropies in intensity and polarization.
More details about the telescope and site can be found in \citet{carlstrom11} and \citet{padin08}.
The primary results in this work use observations from the SPT-3G camera on the SPT.

\subsection{The SPT-3G Camera}
\label{sec:telcam}

The SPT-3G camera replaced the SPTpol camera on the telescope in 2017 and will remain in operation through 2027. 
SPT-3G consists of $\sim$16,000 superconducting detectors configured to observe in three bands, centered at roughly \SI{95}{\giga\hertz} (3.2~mm), \SI{150}{\giga\hertz} (2.0~mm), and \SI{220}{\giga\hertz} (1.4~mm). 
Each camera pixel is coupled to two orthogonally polarized transition edge sensor detectors in each of the three bands.
The band-averaged detector properties relevant to this work are summarized in Table \ref{tab:detector_parameters}.
Effective band centers are calculated for a flat-spectrum and Rayleigh-Jeans spectrum source, showing a modest shift of $\sim$2\%.
For more information on the SPT-3G detectors and readout see also \cite{bender16} and \cite{bender19}.

\begin{table}[ht]
\caption{Adapted from \cite{sobrin22}, the band center is defined as $\int \nu I(\nu)g(\nu) d\nu / \int I(\nu)g(\nu) d\nu$ and band width defined as $\int I(\nu)g(\nu) d\nu$, where $g(\nu)$\ is the normalized SPT-3G frequency response to a beam-filling source, and $I(\nu)$ is the source spectral intensity.
Flat and RJ represent the band center calculated by integrating over a flat and Rayleigh-Jeans spectrum, respectively.
The median noise is per-detector, extrapolated from raw time-ordered data.}

\centering
\begin{tabular}{l c c c}
\toprule
& \textbf{\SI{95}{\giga\hertz}}& \textbf{\SI{150}{\giga\hertz}} & \textbf{\SI{220}{\giga\hertz}} \\
\midrule
Band Center, Flat (\si{\giga\hertz})  & 92 & 145 & 217\\
Band Center, RJ (\si{\giga\hertz})  & 94 & 148 & 221 \\
Band Width (\si{\giga\hertz}) & 26 & 33 & 54\\
Beam FWHM (\si{\arcmin}) & 1.56 & 1.16 & 1.04 \\
Median Noise (mJy $\surd$s) & 26 & 31 & 122 \\ \bottomrule
\end{tabular}
\label{tab:detector_parameters}
\end{table}

The SPT-3G detectors were designed to balance low-noise and dynamic range, with a saturation power of roughly 2$\times$ the estimated optical loading of $\sim$4-8~pW at the South Pole.
The detectors are sampled downstream at a rate of \SI{152.6}{\hertz} and have median noise on short timescales presented in Table \ref{tab:detector_parameters}.
The \SI{220}{\giga\hertz} noise tends to be about 4$\times$ higher than the other two bands, so the SPT-3G \SI{150}{\giga\hertz} band is the most sensitive to sources with a thermal spectrum.
For more details on the SPT-3G instrument and performance, see \cite{anderson18} and \citet{sobrin22}.

\subsection{The SPT-3G Survey}
\label{sec:obs}

One of the benefits of observing from the South Pole is that the celestial sky never rises nor sets.
Because it is located along the axis of rotation of the Earth, the celestial sky simply rotates around the South Pole and declination is equal to elevation at all times.
The main SPT-3G survey region is a 1500 square-degree patch of sky centered at RA~$0^{\text{h}}$, Dec~$-56^\circ$, covering $\pm$\SI{50}{\degree} in RA and $\pm$\SI{15}{\degree} in Dec.
From here on this is called the ``1500d survey".
This field has been observed during the austral winter (March through November) since 2018 with an effective cadence of 1-2 days.
The full field is split equally by elevation into four subfields, two of which are observed during a given cryogenic fridge cycle which is approximately 20 hours long.
Each subfield is observed 2 or 3 times in succession, with each observation taking approximately 2 hours.
The observations are performed by slewing the telescope at constant angular speed in RA, then taking a small step in elevation and repeating, until the full elevation range of the subfield is achieved.
The instantaneous field of view is approximately 2 square degrees, and the telescope scans with a constant speed of 1 degree per second on the azimuth bearing.

The SPT-3G also conducts an extended survey during the austral summer months, when the 1500d survey field becomes sun-contaminated. 
This extended region covers nearly 3000 square degrees, including several sub-fields which get as low in elevation as \SI{28}{\degree}.
These ``summer fields" will therefore host many more satellite passes from high-inclination LEO satellites (see also Section \ref{sec:orbits} and Figure \ref{fig:sat_el_calc}).

\section{Satellites}
\label{sec:satellites}
This section describes the satellites that could potentially impact a survey at the South Pole, with attention to those in low-Earth orbits.
The expected emission mechanisms and relative contributions are calculated for thermal emission, reflected sunlight, and a specific example of cloud-penetrating radar emission.
The received power at Earth and predicted flux density from a generalized satellite are calculated for the case of SPT-3G. 

\subsection{Orbits}
\label{sec:orbits}
Because the SPT is located at the geographic South Pole, only satellites with highly-inclined orbits are visible in the SPT-3G survey region (above \SI{28}{\degree} elevation).
The North American Aerospace Defense Command (NORAD) tracks objects in Earth orbit larger than 10~cm and publishes their orbital information to the SpaceTrack website\footnote[1]{\url{https://space-track.org}} where it can be downloaded in the form of two-line elements (TLEs). 
A query of this database between January 01, 2017 and August 15, 2024 gives a total of approximately 36,000 unique space objects tracked at some point over that time, with roughly 30,000 of those in LEO.
It should be noted explicitly that the number of objects in orbit at any given time is dynamic and that while the recent reentry rate of satellites is roughly 2,000 per year, the number of satellites in orbit continues to grow. 

These objects are also categorized based on their radar cross section (RCS) into three categories; ``Large" ($>1~\text{m}^2$), ``Medium" ($0.1~\text{m}^2 \text{–} 1~\text{m}^2$) and ``Small" ($<0.1~\text{m}^2$).
The use of RCS as a criterion for candidate selection is not optimal (for example, a distinction is not made between an object with a surface area of \SI{100}{\square\meter} or \SI{1}{\square\meter}) but it does serve as a useful proxy for physical size and therefore thermal emission. 
As shown in Section \ref{sec:emission_mechanisms}, only large, nearby satellites need to be considered.\footnote[2]{Nearby does not necessarily mean LEO, as highly elliptical orbits can bring satellites down to a low perigee, sometimes well within the usual LEO altitudes.}

\begin{figure}
\centering
\includegraphics[width=1.0\linewidth]{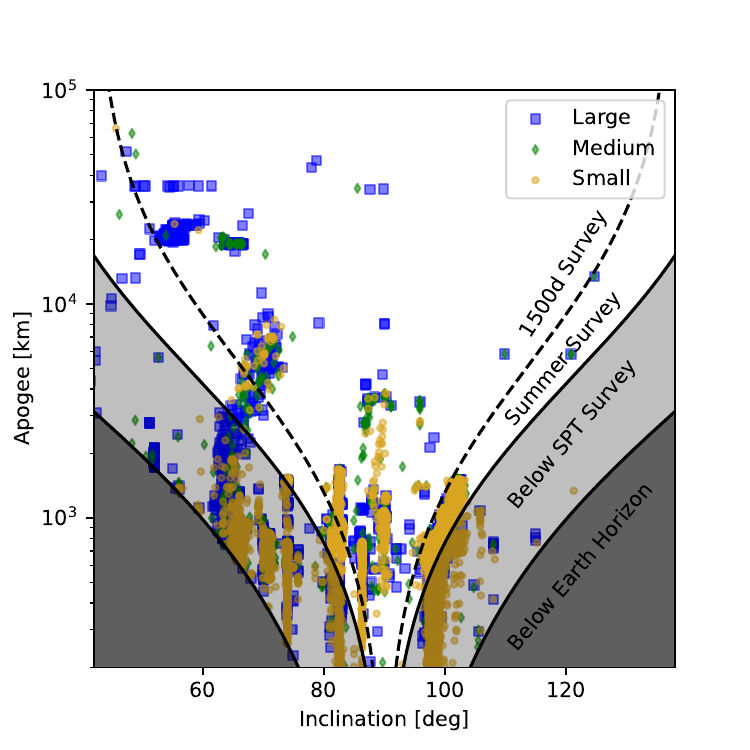}
\caption{\small{Maximum possible elevation of satellites observable from the South Pole, displayed on an inclination-apogee plane. The horizon at the South Pole is designated by the dark shaded region. The light shaded region represents the area above the horizon but below the SPT-3G survey region. Satellites which rise above the survey horizon may be observable. Satellite RCSs are denoted by their markers as shown in the legend.}}
\label{fig:sat_visibility}
\end{figure}

Using the orbital inclination and apogee from the SpaceTrack database, the maximum possible elevation at the South Pole is calculated via simple trigonometry.
Appendix \ref{app:satellite_elevation_calc} describes the geometry and details the calculation of satellite elevation.
A relevant breakdown of all satellites in the SpaceTrack catalog by size is given in Table \ref{tab:satellite_numbers}.
Figure \ref{fig:sat_visibility} shows the apogee-inclination plane for the 21,090 satellites that might rise above the horizon at the South Pole.
Using the apogee provides the upper limit to observed elevation for each satellite without having to construct ephemerides.

\begin{table}
    \caption{Number of satellites, including defunct payloads and debris, broken down by radar cross section that were in orbit between Jan 01, 2017 and Aug 15, 2024. Columns further restrict objects to those in a LEO orbit, above the horizon at the South Pole, and above the SPT-3G survey horizon (El>\SI{28}{\degree}).}
    \label{tab:satellite_numbers}
    \centering
    \begin{tabular}{l c c c c}
    \toprule
       RCS & In Orbit & LEO  & Horizon  & El$>$\SI{28}{\degree} \\
         \midrule
         Large  & 13,571 & 10,578 & 3,691  & 2,286\\
         Medium & 4,958  & 3,879  & 3,022  & 1,534\\
         Small  & 16,136 & 15,218 & 14,377 & 7,957 \\
         \midrule
         Total  & 36,046 & 29,870 & 21,090 & 11,777 \\
         \bottomrule
    \end{tabular}
\end{table}

It is obvious from Figure \ref{fig:sat_visibility} that there are several groupings of satellites into common orbits.
In this paper the focus is placed on LEO, which is defined as an Earth-bound orbit with a period less than 128 minutes.
The corresponding circular orbit has an altitude above the surface of the Earth of approximately 2000~km.
Satellites in LEO will move with apparent angular velocities of order degrees per second and can vary greatly depending on the satellite's altitude and orbital ellipticity.
Of particular note here are the polar orbits which tend to occur at $\pm$\SI{8}{\degree} from the pole due to sun-synchronous considerations.

There are also satellites in elliptical orbits at high inclinations such as the Tundra and Molniya.
These orbits are useful for maintaining long observations over high-latitude regions of Earth and are often inclined at \SI{63.4}{\degree} in order to maintain their apogee above a fixed latitude \citep{kidder90}.
Tundra eccentricities are more moderate ($\sim$0.26) than Molniya ($\sim$0.75) and they both maintain apogees up to some 40000~km (generally over the Northern hemisphere) with perigees as low as a few hundred km in the case of Molniya.
While these orbits are not LEO, they can bring satellites up to an elevation visible from the South Pole, possibly with an altitude of a few hundred kilometers for Molniya or Tundra orbits designed for northern latitude coverage.

Dedicated searches of all visible ``Large" RCS satellites is computationally intensive, thus this work focuses on a few examples of targeted observations and a brief discussion of a blind search pipeline in Section \ref{sec:detecting_satellites}.
It is worth noting explicitly that there are no Geostationary orbits visible above the horizon from the South Pole and that all satellites can be considered point sources to CMB survey experiments (\SI{1}{\arcmin} is $\sim$\SI{300}{\meter} at \SI{1000}{\kilo\meter}).

\subsection{Millimeter-Wave Emission}
\label{sec:emission_mechanisms}

Three contributions to the mm-wave emission are considered.
The first is the intrinsic thermal emission which is radiated isotropically, the second is the reflection of sunlight, and the third is intentional emission from radar or communications systems.
Given the mm-wave bandpass of SPT-3G (1-3~mm), any object that is above $\sim$10~K will radiate on the Rayleigh-Jeans tail of the blackbody curve.
From simple energy balance considerations, LEO satellites are expected to be in the 200-400~K range, which places them safely in the Rayleigh-Jeans regime where the spectral radiance is proportional to $\nu^2$.
No effort is made to estimate the exact value of the emitted flux at any instant since the instantaneous cross-sectional area is not known.
For each observing band, a simple top-hat bandpass is used to calculate the effective emission.

In order to estimate the relative contributions to the total mm-wave emission, the emitted power due to thermal emission and specular reflection are calculated and propagated to the SPT's \SI{10}{\meter} primary mirror.
For simplicity it is assumed that the typical temperature of a satellite is 300~K and that reflected Earthshine and intrinsic thermal emission are indistinguishable.
Detailed calculation of the expected flux due to thermal emission of a satellite is given in Appendix \ref{app:thermal_emission_calc}.
The results are reproduced here, 
\begin{equation}
   S^{\text{therm}}_{\text{150GHz}}(\mathcal{A},d) = 200~\text{mJy} \left(\dfrac{\mathcal{A}}{1~\text{m}^2}\right) \left(\dfrac{1000~\text{km}}{d}\right)^2,\tag{\ref{eq:150GHz_thermal_flux_density}}
\end{equation}
showing, as one would expect, that the flux density is proportional to the effective cross-sectional area, $\mathcal{A}$, and the inverse of the square of the distance, $d$.
It is worth explicitly stating here that $\mathcal{A}$ combines several factors, including emissivity, temperature, and geometric cross-section; therefore even modest uncertainties in these quantities may add up to a large uncertainty in $\mathcal{A}$ and thus the expected thermal emission.

\begin{figure}[h!]
 \centering
 \includegraphics[width=1.0\linewidth]{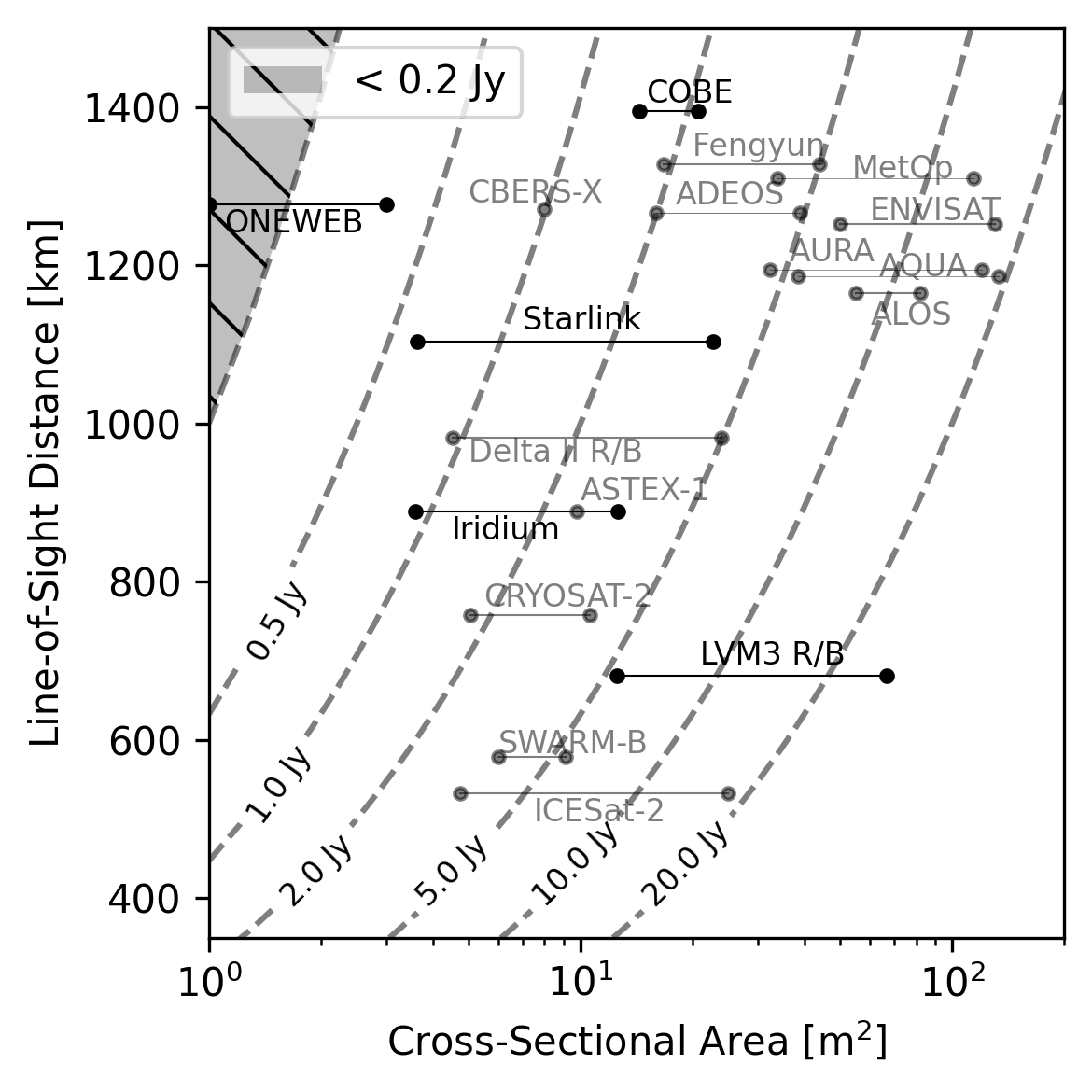}
 \caption{\small{Satellite distance versus cross-sectional area estimated from physical dimensions of the satellite bus and/or solar panels. Lines of constant \SI{150}{\giga\hertz} flux density due to 300~K blackbody emission are drawn as dashed black lines and are calculated using Equation \eqref{eq:150GHz_thermal_flux_density}. The shaded gray region denotes a flux density less than 200~mJy, which is approximately the 1$\sigma$ noise in a well-behaved single bolometer TOD integrated over the $\sim$20~ms that a satellite is expected to be in a single pixel beam. Horizontal lines represent a range that accounts for geometric differences due to satellite bus and/or solar panel orientation. Dark colored text highlights satellites that are discussed as examples in Section \ref{sec:results} of this paper.}}
 \label{fig:sat_diam_vs_LOS}
\end{figure}

Figure \ref{fig:sat_diam_vs_LOS} helps to build intuition for the expected \SI{150}{\giga\hertz} flux density due to thermal emission, calculated in Equation \eqref{eq:150GHz_thermal_flux_density}, for several hand-picked ``Large" RCS LEO satellites which are observable in one of the SPT-3G survey regions.
As an example of a rocket body, a single Delta II stage is shown, which is estimated to be 1/3 of the total length of the rocket.
There are many similar sized rocket bodies in a variety of LEO and non-LEO orbits which make up a large fraction of the current orbital space debris.

To compare the intrinsic thermal emission with reflected sunlight, two extremes are considered.
The first is perfectly Lambertian (or diffuse) reflection and the second is perfectly specular reflection.
The in-band emissivity of a surface, calculated in Appendix \ref{app:thermal_emission_calc} and presented in Table \ref{tab:inband_emittance}, is three orders of magnitude smaller for incident sunlight at 1~AU compared with thermal emission from a 300~K blackbody.
If the reflection is perfectly diffuse it is radiated equally into the hemisphere normal to the surface.
This is exactly the same situation as for thermal emission, which means that under all observing geometries the diffuse reflected sunlight is negligible compared with thermal emission.

In the limit of perfectly specular reflection, however, the reflected light is confined to a diffracted beam.
The specular reflection contributions are calculated in Appendix \ref{app:specular_flare_calc}, including a discussion on the probability of observing such specular flares.
The results take into account both the angular extent of the Sun and diffraction to give 

\begin{equation}
\begin{aligned}
S_\text{150GHz}^{\mathrm{spec}}(A,d) = 3.3~\text{Jy} ~\dfrac{\left ( \dfrac{1000~\text{km}}{d} \right)^2 
                         \left(\dfrac{A}{1~\text{m}^2}\right)}
                         { 0.2 \left ( \dfrac{ 1~\text{m}^2 }{A}\right) + 0.8 } .
\end{aligned} \tag{\ref{eq:Ssun_ref_gen}}
\end{equation}

Comparing the emission from a specular reflection \eqref{eq:Ssun_ref_gen} to the thermal emission \eqref{eq:150GHz_thermal_flux_density} for the nominal case of a 1~m$^2$ blackbody/reflector at 1000~km,
\begin{equation*}    
   \begin{aligned} 
        S^{\text{therm}}_{\text{150GHz}}&(1~\text{m}^2,1000~\text{km}) = 0.2~\text{Jy}\\
        S_{\text{150GHz}}^{\text{spec}}&(1~\text{m}^2,1000~\text{km}) = 3.3~\text{Jy},
    \end{aligned}
\end{equation*}
reveals the surprising result that specular reflections of the Sun are only $\sim15\times$ brighter than the quiescent thermal radiation of a surface of the same size.
In Appendix \ref{app:specular_flare_calc}, the probability of SPT-3G observing specular reflections off of large satellite panels is calculated to be approximately one observed specular reflection per year.

Given the fact that observed satellite flares at optical wavelengths can increase luminosity by many magnitudes \citep{koshkin16, chote19, bassa22}, it is perhaps surprising that at millimeter wavelengths the thermal emission is such a large contributor to the flux.
As discussed in Section \ref{sec:results}, no extreme specular flares have been observed over a total of several hundred observations of ``Large" RCS LEO satellites by SPT-3G, in agreement with the predictions laid out here.

In addition to passive emission, some satellites may also actively transmit at frequencies within SPT's broad observing bands. 
This includes transmissions for communication and remote sensing. 
With active transmissions, the emitted power can be so high that even indirect illumination through a far sidelobe is detectable.
One relevant example of this is the CloudSat satellite, which uses a 1500~W, \SI{94}{\giga\hertz} pulsed radar \citep{spitz01} and has been discussed as a major concern by other radio telescopes including ALMA (see, for example, ALMA Memo No. 504\footnote[3]{\url{https://science.nrao.edu/facilities/alma/aboutALMA/Technology/ALMA\_Memo\_Series/alma504/memo504.pdf}}).
The CloudSat radar is always directed downwards, along the local gravitational vector, thus there is no possibility of SPT to look directly into the main beam because CloudSat neither flies directly over the South Pole, nor does SPT-3G observe the zenith.
In fact, with \SI{98.23}{\degree} inclination and an altitude of 710~km, the CloudSat never gets above \SI{32}{\degree} elevation as viewed from the South Pole.
Even when SPT does observe CloudSat in the main beam, the radar should be approximately \SI{50}{\degree} off-axis.

As shown in Figure 6 of the CloudSat radar characterization paper \citep{durden04}, the far sidelobes are approximately -80~dB.
The 1500~W radar is pulsed in 3.3~$\mu$s pulses at \SI{4000}{\hertz}, thus the resulting effective power is $\sim$20~W in a main beam with 63~dBi.
The effective isotropic radiated power (EIRP) of the main beam is therefore 4$\times10^7$~W and the expected EIRP in the far sidelobe region is 4$\times10^7$~W$\times10^{-8}=0.4$~W. 
Projecting the far sidelobe EIRP to the SPT, the expected power incident on the primary is
\begin{equation}
    P^{cloudsat}_{95\text{GHz}} = P^{cloudsat}_{\text{EIRP}} \dfrac{\pi r^2}{4\pi d^2} = 2.5~\text{pW}~\left(\dfrac{1000~\text{km}}{d}\right)^2.
    \label{eq:cloudsat_power}
\end{equation}
This is roughly 500$\times$ larger than expected power deposited onto the SPT dish from the thermal emission of a 1~m$^2$ object at 1000~km.
The corresponding flux density is 77~Jy, which would make CloudSat the brightest object in any of the SPT-3G survey regions.
For comparison, the brightest astrophysical sources within the 1500d survey field are of order 1~Jy with the brightest peaking around 5~Jy.

A measurement of emission from the far sidelobes of the CloudSat radar is presented in Section \ref{sec:results}, showing good agreement with these calculations.

\section{Detecting Satellites in SPT-3G Data}
\label{sec:detecting_satellites}
This section describes the algorithm for downloading satellite TLE data, generating co-sampled ephemerides, and constructing co-moving maps from which the satellite fluxes are extracted.
While the presentation here is specific to SPT-3G, the algorithm could easily be applied to other observatories that record detector time-ordered data such as the decommissioned Atacama Cosmology Telescope (ACT, \cite{kosowsky03}) or the upcoming Simons Observatory (SO, \cite{ade19}).
In Section \ref{sec:known_satellite_search} the method used for measuring the fluxes of known satellites from SPT-3G data is described.
Section \ref{sec:blind_search} describes a blind detection algorithm which could be used to detect bright satellites without requiring position information.
In Section \ref{sec:results} onward, only the known satellite search results are discussed, with a full blind-search outside the scope of this paper.

\subsection{Known Satellite Search}
\label{sec:known_satellite_search}
In order to extract time-ordered data (TOD) around a known satellite, an ephemeris must be constructed that is sampled at a high-enough cadence such that probable satellite-boresight conjunctions can be calculated.
This is done by first downloading the satellite TLE from SpaceTrack.
TLE data is generated in 6 hour epochs, thus for each SPT observation the nearest-in-time epoch TLE is used.
The Python package \textit{Skyfield}  \citep{skyfield} is then used to generate a 1~second cadence ephemeris for the satellite over the time range of the survey field observation (generally 2 hours for SPT-3G).
If the satellite comes within a conservative distance of the telescope boresight using this 1~second ephemeris, a co-sampled ephemeris is generated within some small time window around the conjunction.
In the case of SPT-3G, co-sampled means taking the time stamps at which each detector is sampled and stored in the TOD, with the sample rate of \SI{152.6}{\hertz}.

Then, for each physical pixel in the focal plane, the satellite-pixel distance at each time sample is calculated.
Pixels which never come within \SI{20}{\arcmin} of the satellite are dropped from the analysis.
For those that are within the distance threshold, the nearest position time sample is calculated and the TOD is excised within $\pm$1~second around that time.
This small time range is chosen to reduce data volume since LEO satellites move with apparent motions of order degrees per second on the sky.
Outside of this window the satellite will have likely left the telescope's field-of-view.
These TOD are then used to make a co-moving map of the satellite by binning the signal in $\Delta$RA, $\Delta$Dec between each detector and the satellite.
Each detector is weighted by the inverse-variance of the TOD outside of the expected position of the satellite so as to not down-weight detectors which see a bright signal.
Due to the limitations of the TLE resolution, model propagation effects, and/or unmodeled maneuvers, the satellites can be quite far from the expected orbits at the time of observation (up to several arcminutes is common, see Section \ref{sec:tle_precision} and Appendix \ref{app:tle_precision}).

Once the TOD are binned into a map, a matched filter is applied which effectively removes any structure larger than the beam and maximizes sensitivity to point sources.
Fluxes are extracted by calculating a weighted map based on the expected thermal-emission sensitivity of each band, calculating the location of the maximum pixel, and using that location to perform forced-photometry in each of the three bands.
The fact that the satellite does not, in general, appear at the expected TLE-derived position (i.e. the center of the co-moving map) makes the extraction of the satellite flux difficult for dim satellites with a brightness near the map noise.
Average flux cannot be simply calculated by stacking maps from multiple observations, for example.
Quantifying and understanding the ephemeris uncertainty is the focus of future work, but is outside the scope of this paper.

SPT-3G fluxes derived using this mapmaking method have been calibrated by running the pipeline on known, bright sources.
Absolute calibration has been tied to ALMA calibrator sources, although there have been few coincident observations at ALMA \SI{155.0}{\giga\hertz} / SPT \SI{150}{\giga\hertz} or at ALMA \SI{233.0}{\giga\hertz} / SPT \SI{220}{\giga\hertz}.
ALMA absolute calibration uncertainties are generally quoted at 5\% in these bands and an internal SPT/ALMA calibration check show up to 5-10\% variations for single observations.
Combining these conservatively, a single SPT-3G measurement may have a flux calibration uncertainty of approximately 10\%.
The total integration time in these maps is challenging to quantify in general, but rough approximations are that each detector observes a satellite for a few milliseconds, with between 10-100 detectors per band contributing in a single pass, yielding a total integration time on the order of tens to hundreds of milliseconds.

\subsection{Blind Satellite Search}
\label{sec:blind_search}
Searching over all known satellites and generating ephemerides for each two~hour observation with a 1-second cadence is computationally intensive and inefficient.
One could imagine instead searching the time ordered data for moving object tracks.
Because LEO satellites move of order degrees per second on the sky, they will only be within the focal plane for a maximum of several seconds, depending on the relative motion between the satellite and the telescope scan direction.
This analysis requires a new method of data analysis which extracts the raw TOD before many of the common filtering steps such as ``glitch finding" which would flag a bright satellite as a bolometer glitch.

To begin, the raw TOD are first calibrated into on-sky temperature units before being filtered using a Ricker wavelet, which is Gaussian in shape but has zero mean. 
This is an extremely harsh filter that removes all long-timescale fluctuations, effectively removing everything larger than the beam size.
The filtered TOD are then analyzed using a sliding time window.
The RMS of the region outside of the sliding time window is calculated and used to define significant outliers within the sliding window.
Every time a significant extremum is found, it is masked, the window is moved, and the process continues, recording each significant extremum along the TOD.

\begin{figure}[h!]
\centering
\includegraphics[width=1.0\linewidth]{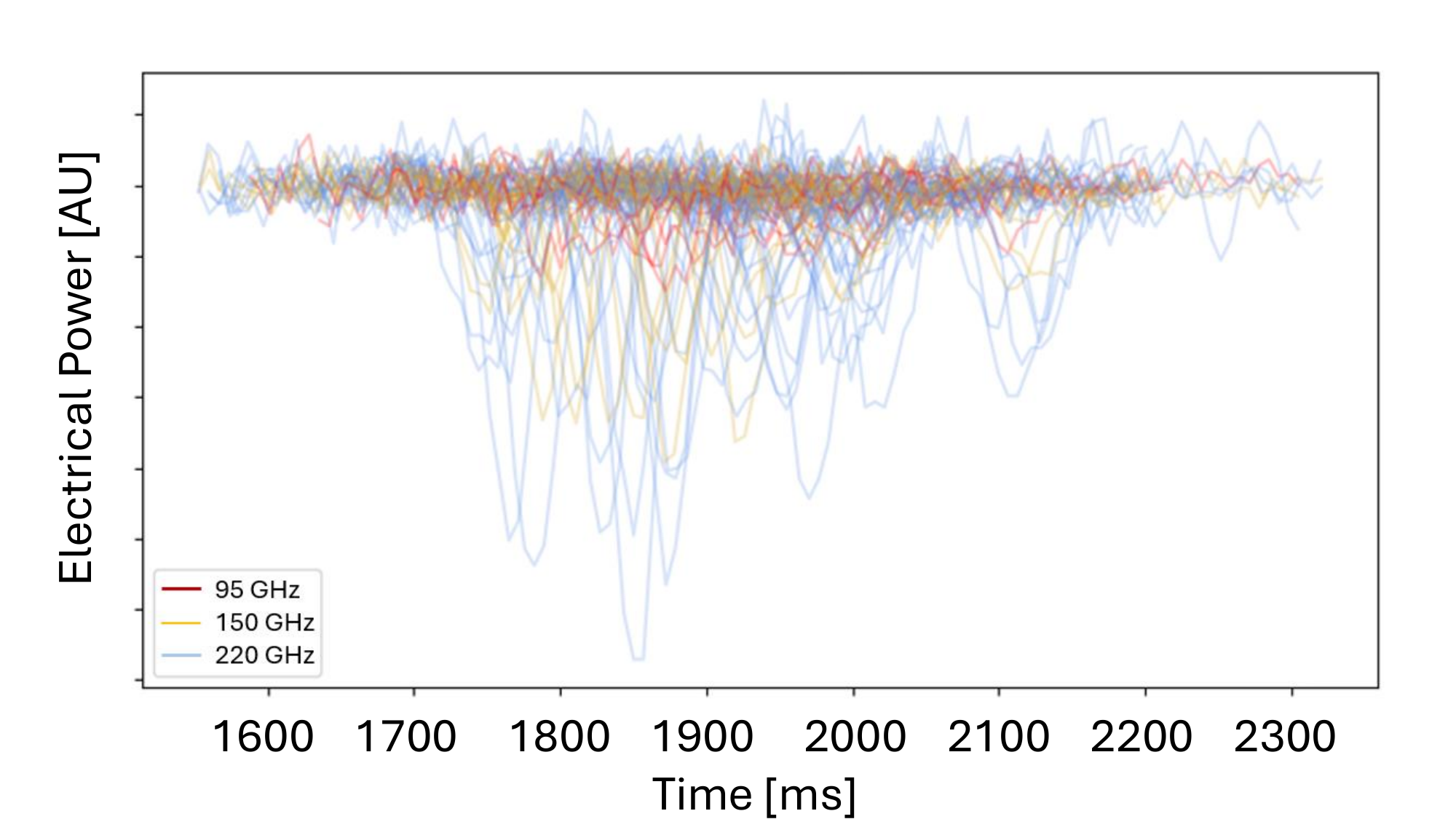}
\caption{\small{The time ordered data of high signal-to-noise bolometers for a single event found using the blind satellite search. Power is shown in arbitrary units and line colors differentiate detector observing band. The time is relative to the start of the sliding time window used in the search.}}
\label{fig:sat_tod_example}
\end{figure}

After the TOD from all bolometers are analyzed in this manner, temporally connected bolometer events are grouped by binning the extrema time.
An astrophysical source on the sky is within the telescope's field-of-view for a length of time given roughly by the focal plane size divided by the scan speed.
Thus for SPT-3G, an object will be visible for roughly 2 seconds (2~deg / 1~deg~s$^{-1}$), and therefore a moving object like a satellite can be taken to be visible for a similar amount of time; shorter if the satellite is moving against the telescope motion and vice versa.
Time bins are therefore set to 500 samples which equates to nearly 3.3~seconds.
Bins containing a significant number of bolometer-events are flagged for further analysis.
An example of the TOD for a single flagged event is shown in Figure \ref{fig:sat_tod_example}.

\begin{figure}[h!]
\centering
\includegraphics[width=1.0\linewidth]{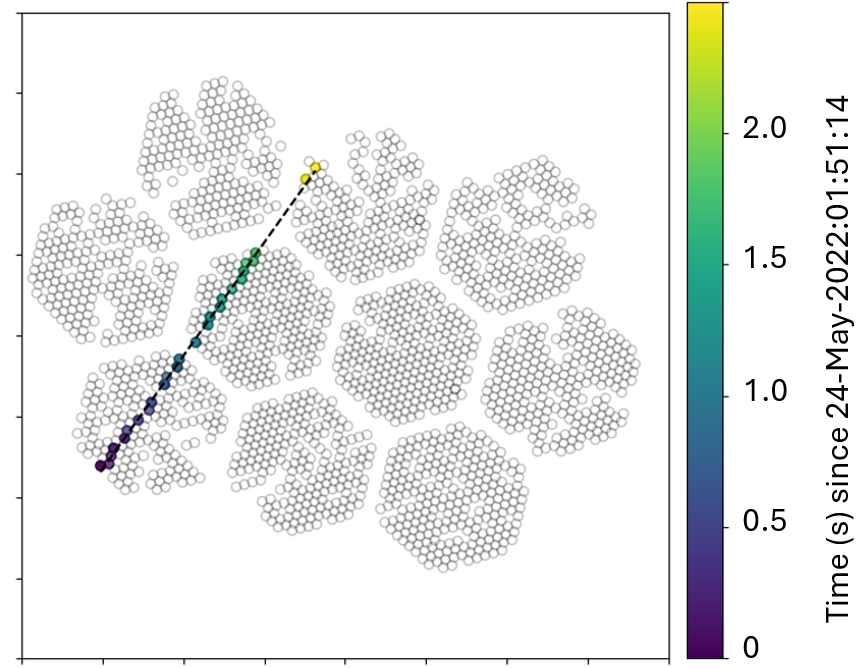}
\caption{\small{A satellite track, defined by the timing of detections by high signal-to-noise bolometers, projected on the SPT-3G focal plane. Individual pixels are shown as black circles, with each pixel nominally containing six detectors (two polarizations in each of the three observing bands). Some pixels are missing due to readout issues. The black dashed line shows the best-fit line and pixel colors correspond to the relative time of detection in seconds.}}
\label{fig:sat_fp_example}
\end{figure}

Each flagged bin is then analyzed to look for linear motion across the focal plane.
The width of the time bins (sliding window size) is important to fine-tune such that any bin might contain one, but no more, satellites.
Allowing for multiple satellites in one time-bin would require a more sophisticated algorithm for differentiation of multiple linear features.
This has not been implemented because in the SPT-3G survey area the satellite density is rather low and conjunctions of two or more satellites at the same time are unlikely (see Section \ref{sec:effects_on_cmb_surveys} for more details).
Figure \ref{fig:sat_fp_example} shows the linear feature detection for the event shown in Figure \ref{fig:sat_tod_example}, with the line-of-best-fit drawn over-top.

\begin{figure}[h!]
\centering
\includegraphics[width=1.0\linewidth]{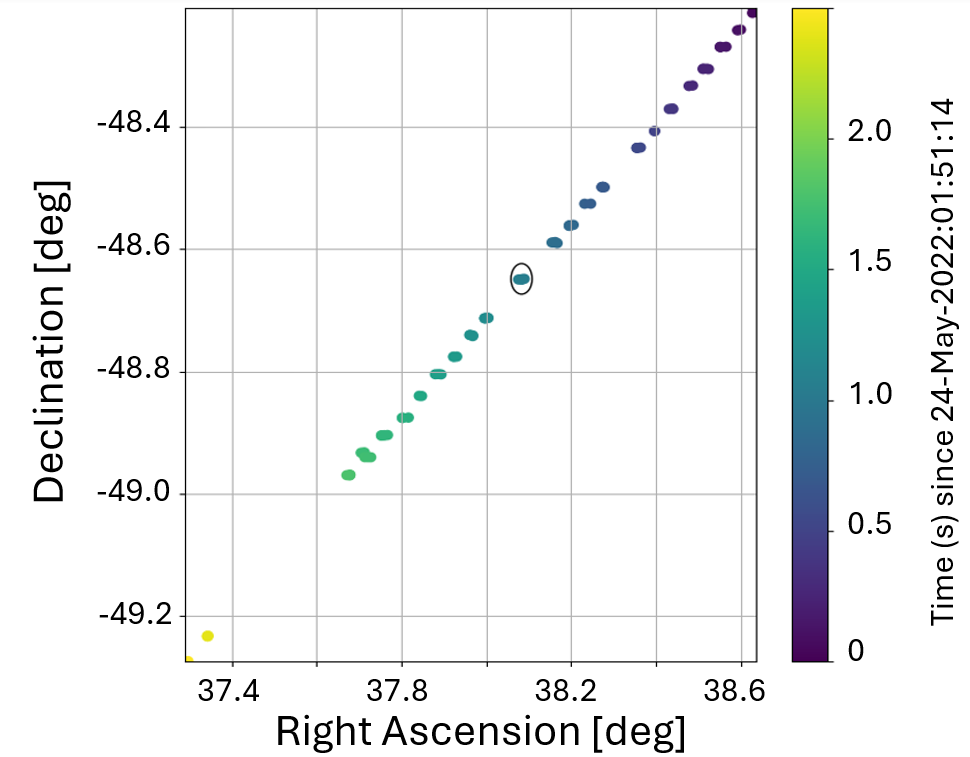}
\caption{\small{Motion of the example satellite after subtracting the SPT focal plane motion. The black circle is 3$'$ diameter around the mean position and the marker colors correspond to the relative time of detection in seconds. A stationary source would show a scatter of points inside the black circle.}}
\label{fig:sat_sky_example}
\end{figure}

Removing events for which the $\chi^2$ of a fit to a time-ordered line across the focal plane exceeds a certain value cuts out most of the undesired signals including readout-related noise, cosmic rays, or small-scale, high-amplitude atmospheric fluctuations.
Once linear motion events are detected, the focal-plane motion can be subtracted leaving only the residual on-sky motion. 
Astrophysical sources thus become a smattering of points within the angular diameter of the telescope's beam and can be easily matched to known, cataloged sources.
Remaining moving sources can then be analyzed using the recorded high signal-to-noise bolometers, on-sky positions, and times of detection.
Because the density of moving objects over the South Pole is relatively low, the mean position and time (i.e. the value within the black circle of Figure \ref{fig:sat_sky_example}) can be used with a generous padding to perform a cross-match with known satellites.
A query of known LEO satellite positions within 2~degrees of the mean position and time of the example shown here returns one satellite: ICESAT-2 (NORAD ID 43613).

\section{Results}
\label{sec:results}

In a novel, targeted search, SPT-3G has observed several ``Large" RCS LEO satellites with evidence for the measurement of thermal emission as well as intentional in-band emission from the CloudSat \SI{94}{\giga\hertz} radar.
These satellites can be several Jy at \SI{150}{\giga\hertz} and individual satellites can vary in observed flux by factors of several depending on orientation, line-of-sight distance, and satellite temperature.
The measured fluxes of satellites observed in sunlight and in the shadow of the Earth are consistent, confirming that observing solar specular reflections is extremely rare, and that, in general, they are sub-dominant to thermal radiation. 
Similarly, searches for specular flares from several hundred observations of satellites in OneWeb, Iridium, and Starlink constellations did not turn up any significant flares (i.e. the measured fluxes were all consistent with the blackbody expectation).

Observations from four months of the SPT-3G survey were analyzed in this work, June-July 2022 and January-February 2023, which cover two austral winter and two austral summer months.
The summer observations were chosen to include some low-elevation survey regions which allow for the measurement of Starlink satellites in polar orbits.
While there are 2,286 ``Large" RCS LEO satellites above the SPT survey horizon at the time of writing of this paper, only a few are hand-selected to be shown as examples here.
This is due in large part to the computational requirements to search for large numbers of satellites and the coarse definition of ``Large" RCS which does not guarantee that the source is bright enough to be observed in a single SPT-3G observation.
In particular, OneWeb satellites are labeled ``Large" in the SpaceTrack catalog, but are below the single-observation noise level in each SPT-3G observing band.
Of nearly 1,000 OneWeb conjunctions over these four months, there were no significant detections.

To begin this section, the results of two particular thermal-emission case-studies are presented.
The first is a decommissioned CMB survey satellite (COBE) and the second is a set of cylindrical rocket bodies.
Building on these examples, observations of two well-known satellite constellations, Iridium and Starlink, are discussed.
The final result is the observation of non-thermal emission from an actively emitting, Earth-observing radar.
Additionally, a systematic offset between the observed positions of these satellites and their expected locations based on TLE predictions is reported. 
The discrepancies suggest either inaccuracies in the TLE data or unmodeled orbital perturbations, both of which warrant further investigation.

\subsection{The Cosmic Background Explorer (COBE)}
\label{sec:COBE}

The Cosmic Background Explorer (COBE) satellite was a mission operated by NASA from 1989 to 1993, dedicated to studying the frequency spectrum and angular distribution of the CMB.
Data from COBE was used to demonstrate that the universe was consistent with a blackbody at a temperature of 2.7~K to high precision \citep{mathur90,fixsen96} and to detect that the small anisotropies in this temperature are on the order of 1 part in 10$^5$ \citep{smoot92}.
For these discoveries, the COBE team was awarded the Nobel Prize in physics in 2006.
Since 1993, however, COBE has been floating as space debris in its nominal sun-synchronous orbit with inclination \SI{99}{\degree} and altitude 900~km.
As an ode to this pioneering instrument, it is included in this work as an example for the detection of thermal emission.

Using the drawing and physical dimensions found in \cite{boggess92},  the cross-sectional area of the COBE bus is estimated to be 17.4~m$^2$ and the solar panels to be 8.3~m$\times$2.5~m = 20.8~m$^2$.
Depending on the observed angle, the cross-sectional area could be the sum of the bus+panels or simply the end-on area of the conical Sun/Earth shield (roughly 20~m$^2$).
Since the satellite is decommissioned (and has been for more than 30 years), the orientation is no longer being tracked. 
The cross-sectional area is therefore taken to be an average of the minimum and maximum cross-sectional area described above.
This gives an effective cross-section of $\mathcal{A}$=23.5~m$^2$.

\begin{figure}[h!]
\centering
\includegraphics[width=1.\linewidth]{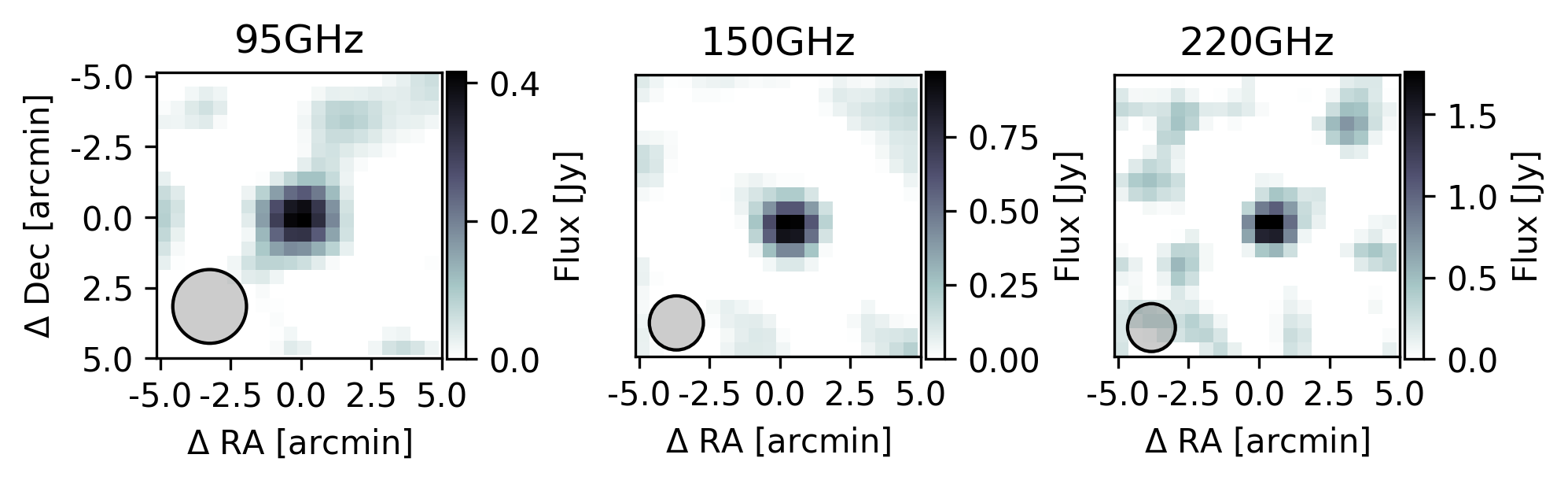}
\caption{Single observation, co-moving map centered on the measured position of COBE in each of the three SPT-3G observing bands. The increasing flux density with increasing frequency is consistent with thermal emission, as stated in the text. The SPT-3G beam is shown as a shaded 2$\sigma$ contour in the lower-left.}
\label{fig:cobe_maps}
\end{figure}

SPT-3G observed a COBE pass on January 20th, 2023 at a distance of 1440~km.
If one assumes a 300~K blackbody, the predicted flux density from Equation \eqref{eq:150GHz_thermal_flux_density} is 2.2~Jy at \SI{150}{\giga\hertz}.
Calculating the same for the other two frequency bands gives 1~Jy at \SI{95}{\giga\hertz} and 5~Jy at \SI{220}{\giga\hertz}.
SPT-3G measurements give flux densities of 0.4$\pm$0.1~Jy, 1.0$\pm$0.1~Jy, and 1.8$\pm$0.2~Jy, at 95, 150, and \SI{220}{\giga\hertz}, respectively, as shown in Figure \ref{fig:cobe_maps}.  
Although these appear to be discrepant by a factor of two, COBE is clearly not a perfect blackbody (intentionally so, as NASA used an aluminized thermal shield to reflect radiation from the Sun and Earth).
If an emissivity of 0.04 is assumed, the expected fluxes are 0.5, 1.2, 2.7~Jy at 95, 150, and \SI{220}{\giga\hertz} respectively.
Furthermore, it is difficult to estimate the cross-sectional area observed on any given pass.
With all of these unknowns, the SPT-3G measurements are well within the limits calculated using reasonable approximations.

\subsection{Cylindrical Rocket Bodies}
\label{sec:lvm3_thermal_emission}
\begin{figure}[h!]
\centering
\includegraphics[width=1.\linewidth]{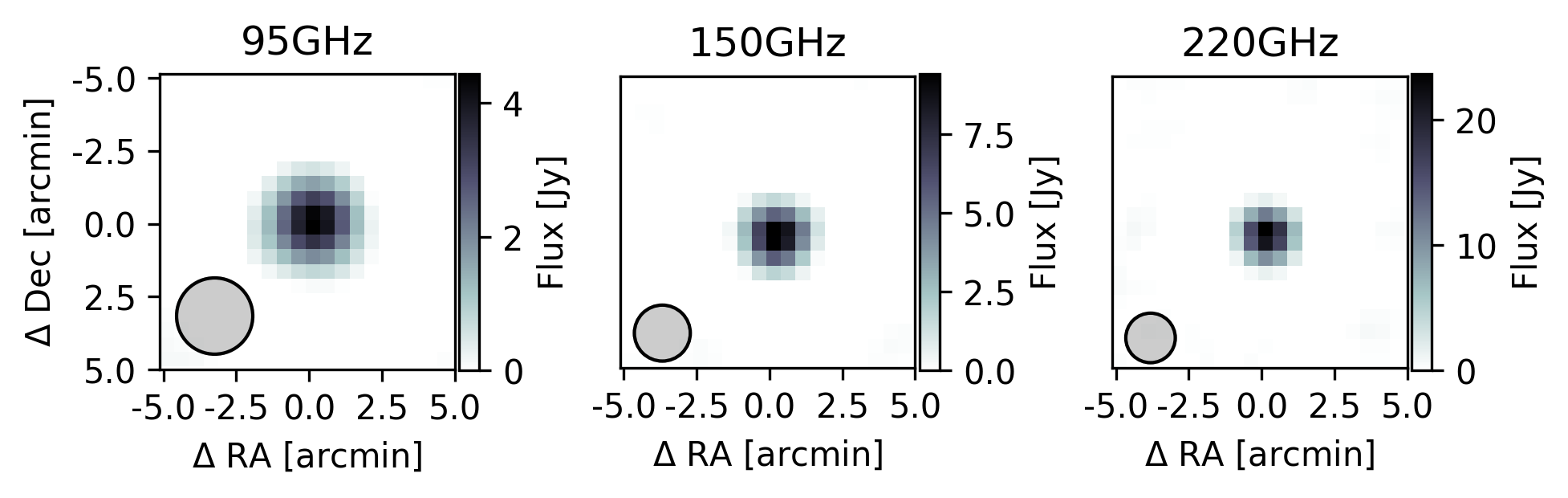}
\caption{Example of a single observation, co-moving map centered on the measured position of a LVM3 rocket body in each of the three SPT-3G observing bands. The SPT-3G beam is shown as a shaded 2$\sigma$ contour in the lower-left.}
\label{fig:lvm3_maps}
\end{figure}

To better constrain the expected thermal emission, cylindrical rocket bodies can be studied because they are not actively emitting and they have simple geometries.
Two prime examples are upper-stage rocket bodies of the Indian Space Research Organisation (ISRO) Mark III launch vehicles (LVM3s) which were launched in late 2022 and early 2023 into polar low-Earth orbits (NORAD IDs 54149 and 56082).
These 4~m diameter, 13.5~m tall cylinders provide an easily calculable minimum and maximum surface area for passive emission \citep{isrolvm3}.
Figure \ref{fig:lvm3_maps} shows an example of a single SPT-3G co-moving map of a LVM3 rocket body. 

\begin{figure}[h!]
\centering
\includegraphics[width=1.\linewidth]{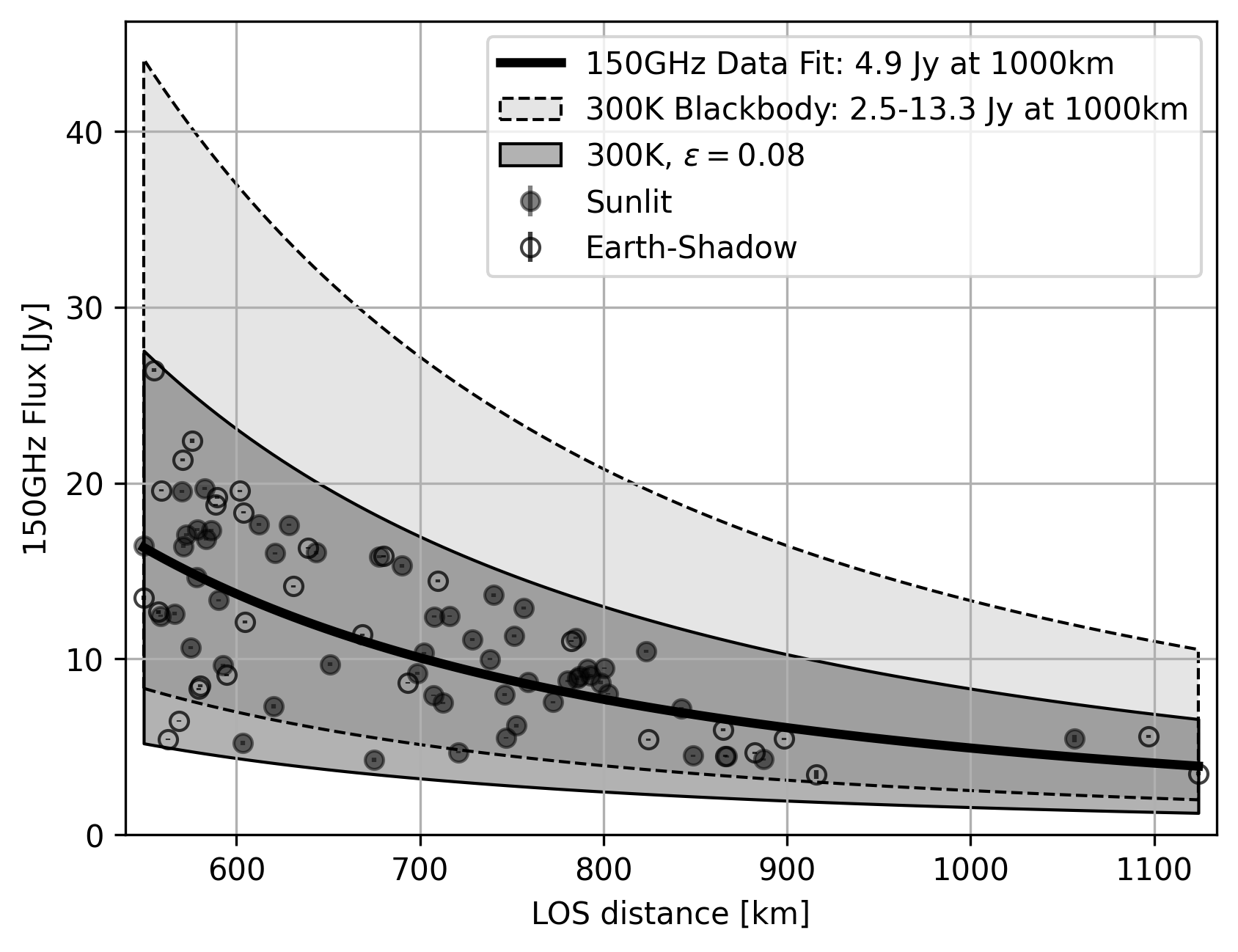}
\caption{Scatter plot of measured fluxes versus line-of-sight distance for two LVM3 upper stages (NORAD IDs 54149 and 56082). Sunlit and shadowed observations are denoted by filled and empty markers, respectively. The heavy line shows the d$^{-2}$ best-fit power law. The light shaded region shows the flux limits expected for a 300~K cylindrical blackbody with dimensions 4$\times$13.5~m as a function of line-of-sight distance. A best-fit emissivity of 0.08 is shown as the darker shaded region.}
\label{fig:gslv_flux_vs_distance}
\end{figure}

The \textit{skyfield} package \citep{skyfield} is used to calculate whether or not the object was illuminated by sunlight during a given observation.
Out of 140 observations, there were 60 which occurred while the satellite was in the Earth's shadow.
After making a co-moving map for each observation, the flux and noise are extracted; the results for the LVM3 rocket bodies are shown as a function of the line-of-sight distance in Figure \ref{fig:gslv_flux_vs_distance}.
The measured fluxes are consistent between the sunlit and shadowed distributions with the 90$^{\text{th}}$-percentile flux uncertainty approximately 0.1~Jy at 150~GHz.
While the LVM3 rocket body temperature and millimeter emissivity are not known, the data shown in Figure \ref{fig:gslv_flux_vs_distance} suggest that a 300~K blackbody alone may not be the best fit.
Using the best-fit to the measured fluxes projected onto a fiducial distance of 1000~km (4.9~Jy at \SI{150}{\giga\hertz}) and an average geometric cross section of a randomly tumbling cylinder (41.5~m$^2$ for the LVM3 upper stage,\citealt{klett64}), a best-fit emissivity is calculated to be 0.08 as shown by the darker shaded region in Figure \ref{fig:gslv_flux_vs_distance}.
This is a reasonable emissivity for an aluminum alloy, however, other degenerate factors such as temperature and effective cross-sectional area could play a role.

\begin{figure}[h!]
    \centering
    \includegraphics[width=1\linewidth]{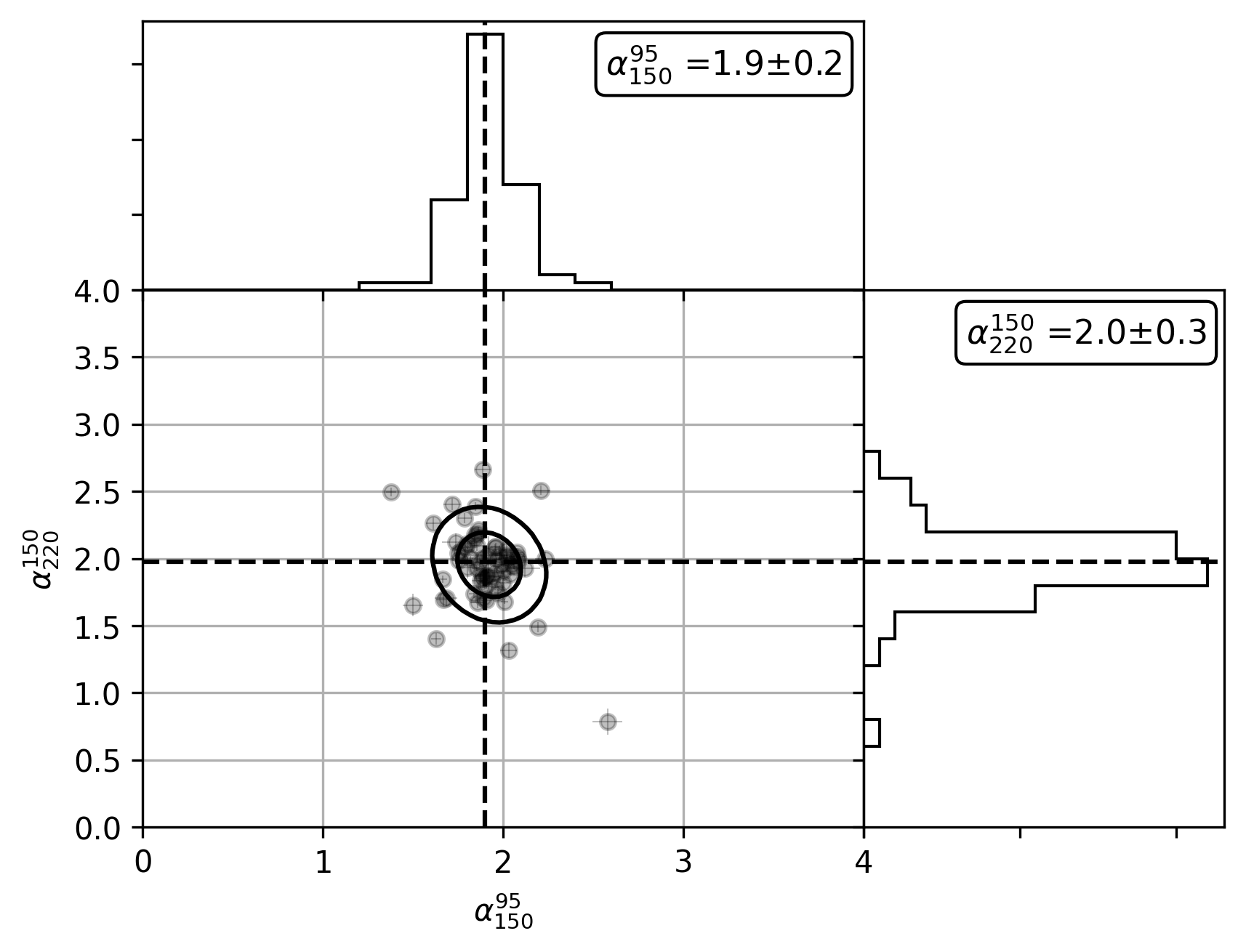}
    \caption{Spectral index distribution for the two LVM3 rocket bodies. A 2D Gaussian is fit to the data and shown as 1- and 2$\sigma$ contours over the scatter plot. Histograms of the two-band spectral indices are shown on the sides of the plot with black dashed lines indicating the inverse-variance (IV) weighted averages along each axis. The legends show the IV-weighted average and standard deviation.}
    \label{fig:gslv_spectral_index_scatter_w_histogram}
\end{figure}

The three observing bands are used to calculate the spectral index, $\alpha$, of the emission.
Defining $\alpha$ such that 
\begin{equation}
    S_{\nu} \propto \nu^{\alpha},
\end{equation}
i.e.
\begin{equation}
    \alpha = \dfrac{\text{log}(S_{\nu_1} / S_{\nu_2})}{\text{log}(\nu_1/\nu_2)},
\end{equation}
$\alpha$ is calculated using the two pairs of bands, namely, 95~GHz/150~GHz and 150~GHz/220~GHz.
The resulting measurements and their standard deviations are shown in Figure \ref{fig:gslv_spectral_index_scatter_w_histogram}.
The mean and standard error (std. err.) of the 95~GHz/150~GHz spectral index ($\alpha^{95}_{150}$) is equal to 1.90$\pm$0.03, in slight tension with the expected $\nu^2$ dependence of thermal emission.
The higher frequency $\alpha^{150}_{220}$~=~1.98 with a std. err. of 0.04, consistent with thermal emission.

\subsection{Starlink Constellation}

\begin{figure}[h!]
\centering
\includegraphics[width=1.\linewidth]{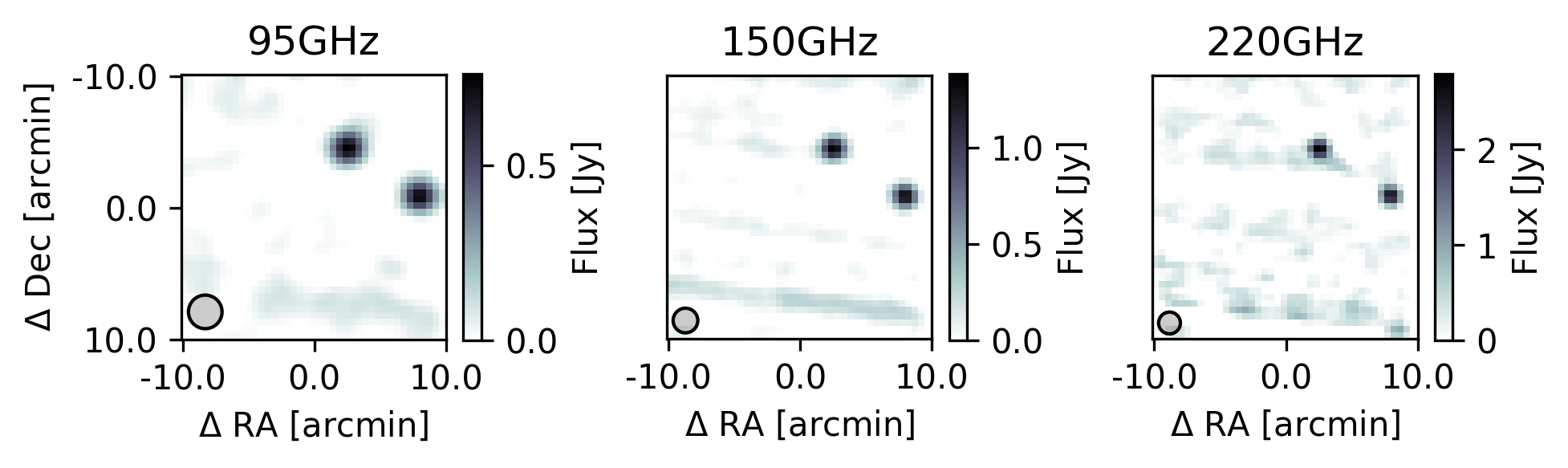}
\caption{Co-moving maps centered on the nominal, TLE-derived position of Starlink 4377. The other satellite in the map is Starlink 4311, which happened to be nearby and on a similar orbit. The SPT-3G beam is shown as a shaded 2$\sigma$ contour in the lower-left.}
\label{fig:starlink_maps_double}
\end{figure}

Although Starlink satellites do not enter the SPT-3G 1500d survey region, they do appear in the extended survey region which reaches as low as \SI{28}{\degree} in elevation.
A low-elevation subfield was observed in February 2023, recording several conjunctions with satellites in the Starlink constellation.
In fact, the spatial density was high enough that two Starlink satellites (4377 and 4311) were observed in a single \SI{20}{\arcmin}$\times$\SI{20}{\arcmin} map, as shown in Figure \ref{fig:starlink_maps_double}).

\begin{figure}[h!]
    \centering
    \includegraphics[width=1\linewidth]{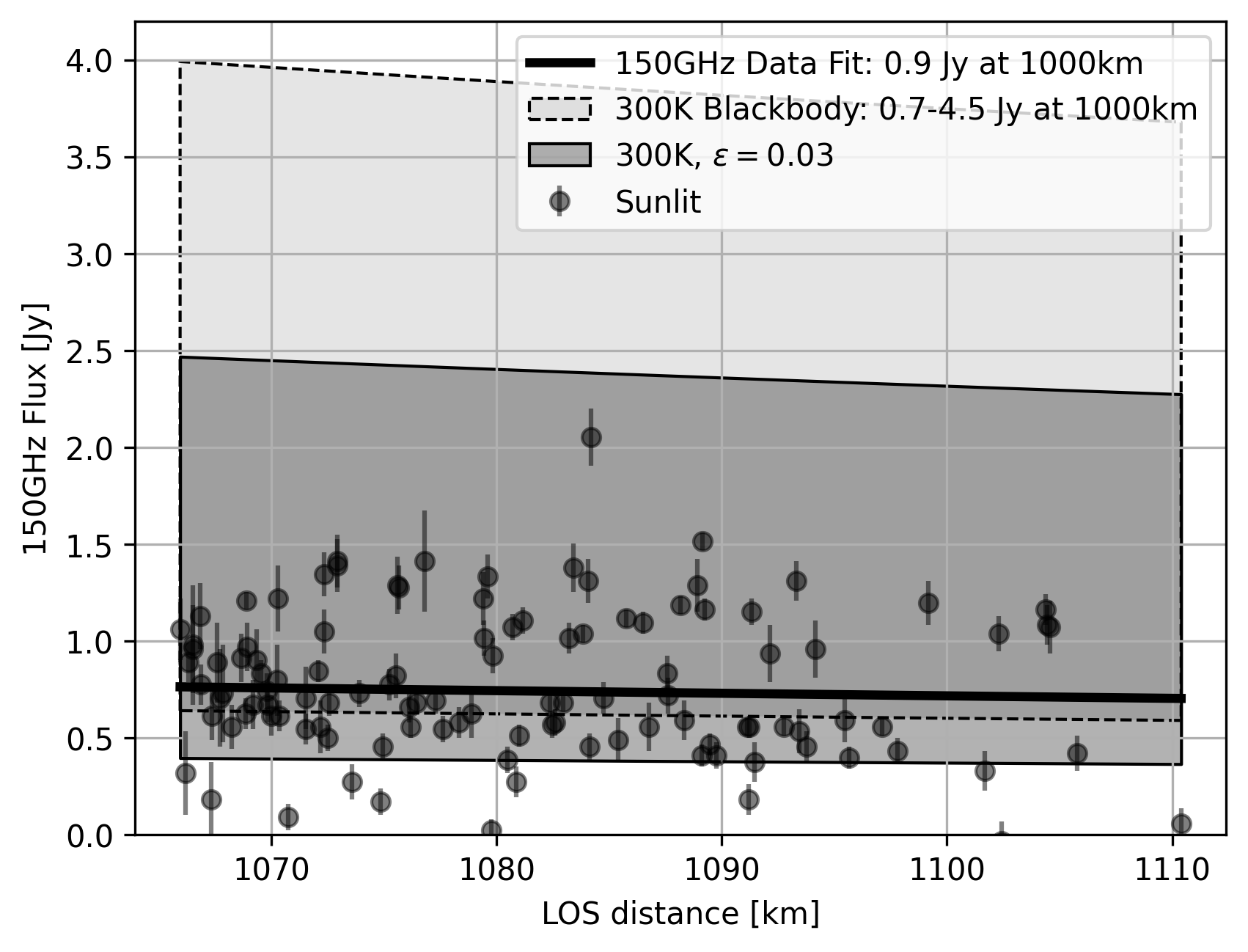}
    \caption{Same as Figure \ref{fig:gslv_flux_vs_distance} but for Starlink satellites. The measured fluxes are consistently on the lower side of the 300~K blackbody range. Observations took place during austral summer so all measurements were in sunlit conditions. The dark shaded region assumes a polished aluminum emissivity of $\varepsilon=0.03$.}
    \label{fig:starlink_flux_vs_distance}
\end{figure}

Starlink Gen2 satellite filings state the bus size is 1.3$\times$2.8~m with a solar array that is 2.8$\times$8.1~m \citep{goldman22}.
This gives a large range in expected fluxes, especially since the form factor at any given observation is unknown a priori.
At a fiducial distance of 1000~km, the expected \SI{150}{\giga\hertz} flux from Starlink satellites ranges from 0.7~Jy to 4.5~Jy assuming a 300~K blackbody.
The measured SPT-3G \SI{150}{\giga\hertz} fluxes have a projected mean of 0.9$\pm$0.4~Jy at 1000~km, which are consistent with the expectations, albeit at the low end of the expected flux range (see Figure \ref{fig:starlink_flux_vs_distance}).
Even when assuming a polished aluminum emissivity of $\varepsilon=0.03$, the measurements lie at the lower end of the expectation, indicating that the estimated upper-end of the cross-sectional area is not consistent (i.e. one does not observe the solar panel and bus both face-on simultaneously).
The 90$^{\text{th}}$-percentile flux uncertainty for these observations is 0.15~Jy at 150~GHz.
Because these observations took place in February, the Sun is always above the horizon at the South Pole and thus there are no shadowed observations.

Although the scatter in the Starlink spectral index measurements is larger than for the bright LVM3 rocket bodies described in Section \ref{sec:lvm3_thermal_emission}, the mean and std. err. are measured to be $\alpha^{95}_{150}=1.41\pm0.06$ and $\alpha^{150}_{220}=2.09\pm0.07$.
Again this shows that the higher frequency measurements are more consistent with thermal emission and that there is a larger tension measured between 95~GHz and 150~GHz.
It is not fully understood why the measurements of the Starlink $\alpha^{95}_{150}$ are low, implying a $\sim$30\% flux excess in the SPT-3G 95~GHz band.

\begin{figure}[h!]
    \centering
    \includegraphics[width=1\linewidth]{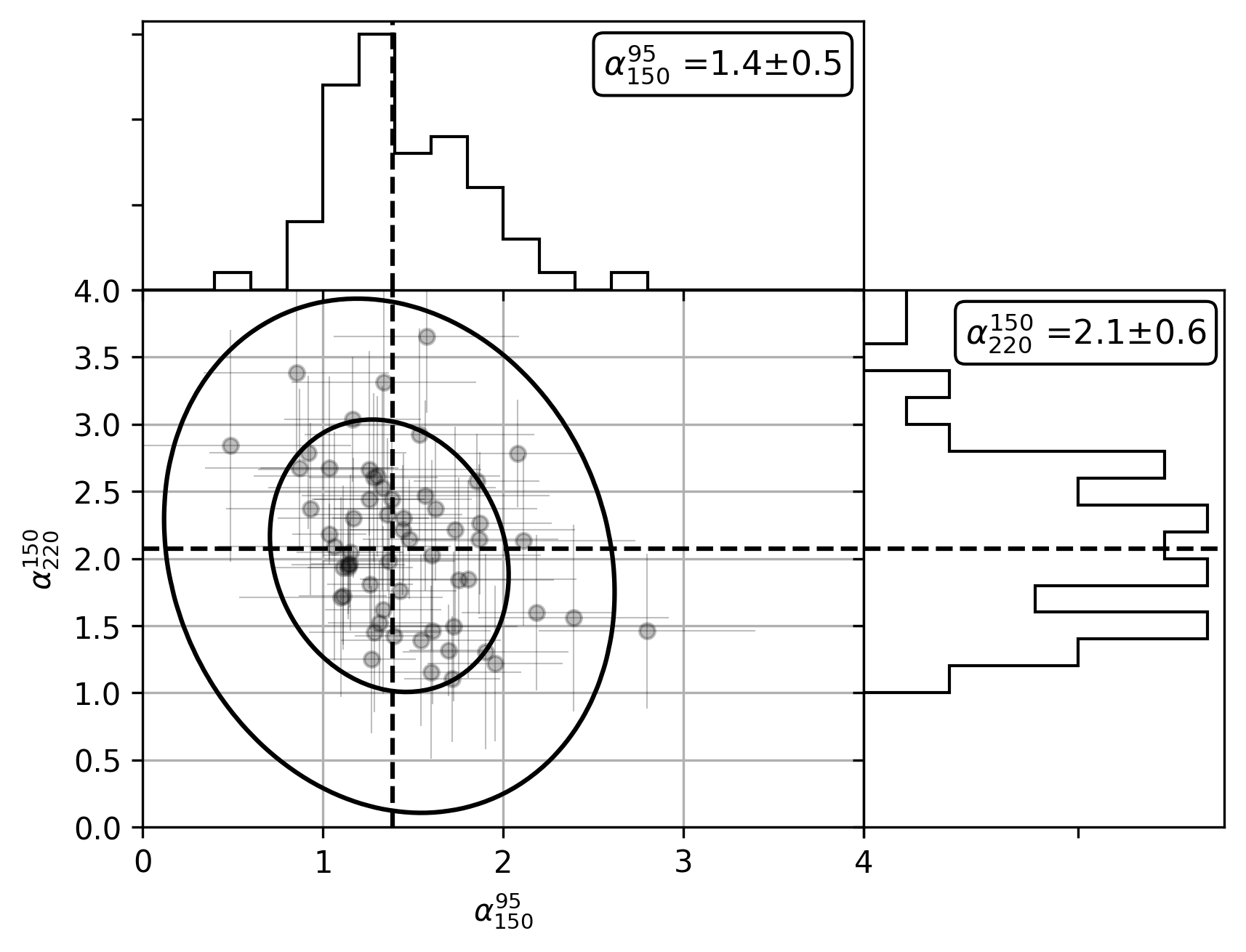}
    \caption{Spectral index distribution for the observed Starlink satellites. A 2D Gaussian is fit to the data and shown as 1- and 2$\sigma$ contours over the scatter plot. Histograms of the two-band spectral indices are shown on the sides of the plot with black dashed lines indicating the inverse-variance (IV) weighted averages along each axis. The legends show the IV-weighted average and standard deviation.}
    \label{fig:starlink_spectral_index_scatter_w_histogram}
\end{figure}

\newpage
\subsection{Iridium Constellation}

The Iridium satellites comprise a well known constellation, the first generation of which produced extremely bright optical flares from specular reflection off of their three, \SI{120}{\degree} separated antennae.
While this is true in the optical, no specular flares have been measured in any of the 225 observations of Iridium satellites in this work.
The Iridium satellites have three antennae, with dimensions of 1.5$\times$2.4~m, that are angled at \SI{40}{\degree} with respect to the satellite bus which is oriented along nadir when in controlled orbit. 
Each satellite also has a solar array of size 1.5$\times$8.4~m \citep{maley03}.

\begin{figure}
    \centering
    \includegraphics[width=1\linewidth]{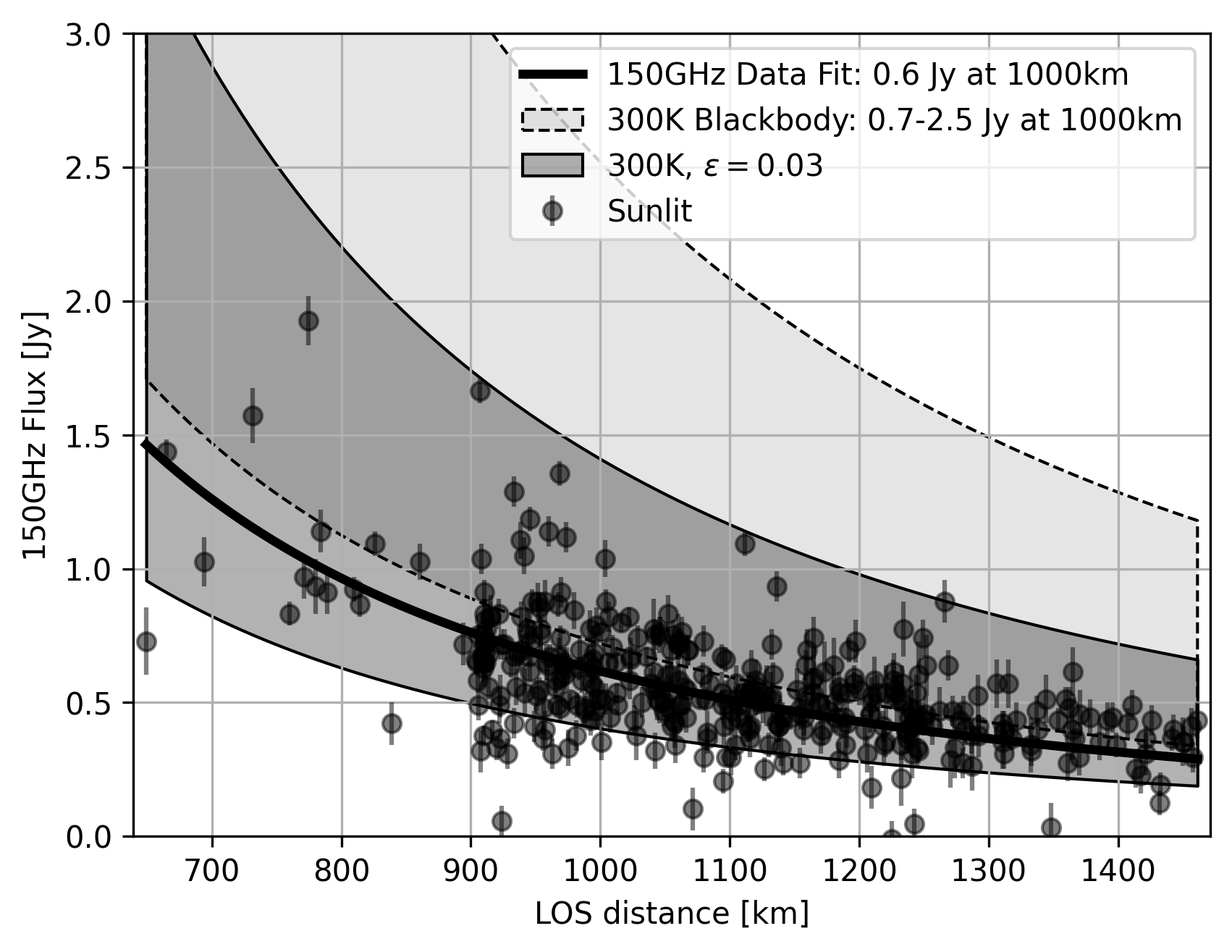}
    \caption{Same as Figure \ref{fig:starlink_flux_vs_distance} but for Iridium satellites. Observations took place during austral summer so all measurements were in sunlit conditions.}
    \label{fig:iridium_flux_vs_los_dist}
\end{figure}

\begin{figure}
    \centering
    \includegraphics[width=1\linewidth]{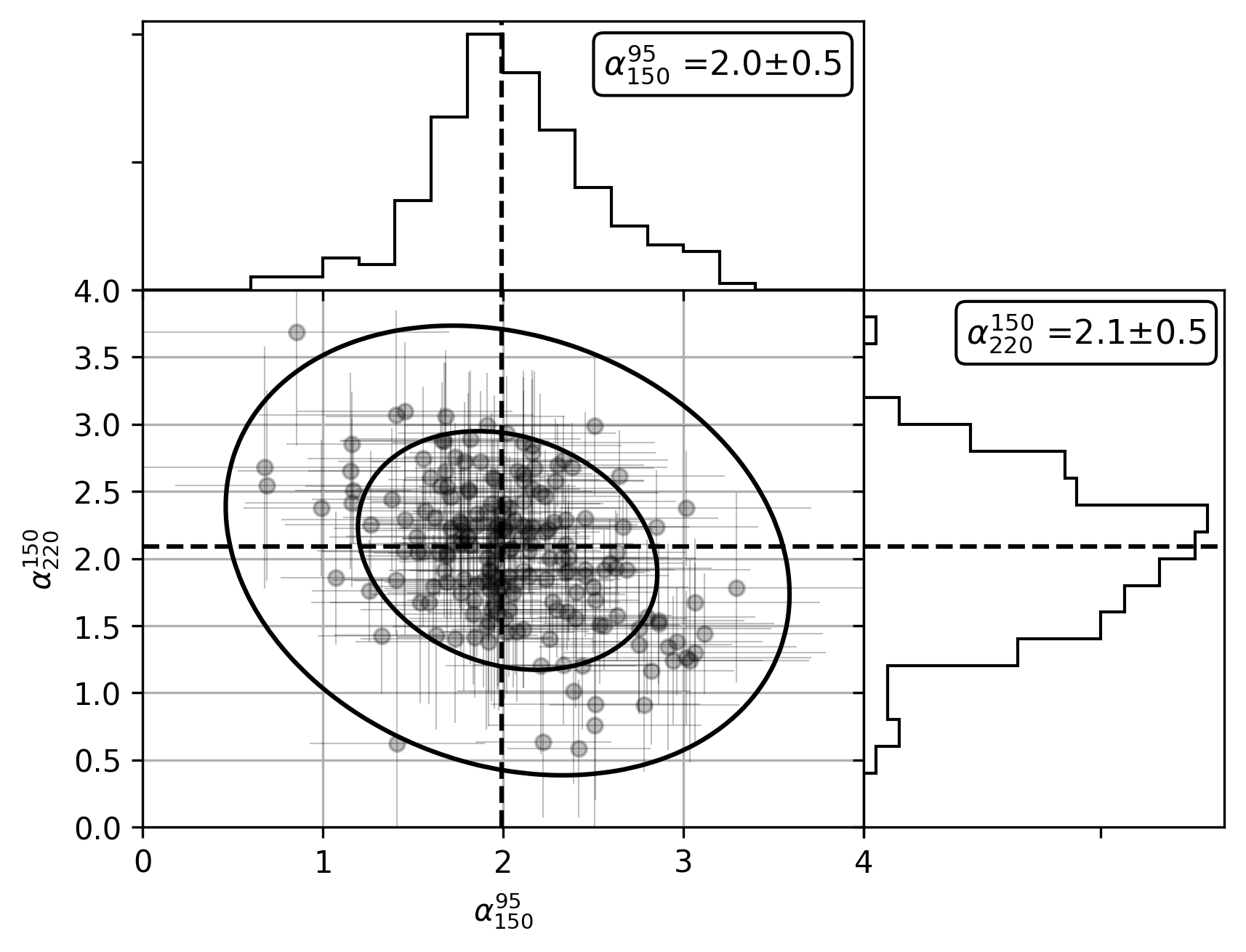}
    \caption{Spectral index distribution for the observed Iridium satellites. A 2D Gaussian is fit to the data and shown as 1- and 2$\sigma$ contours over the scatter plot. Histograms of the two-band spectral indices are shown on the sides of the plot with black dashed lines indicating the inverse-variance (IV) weighted averages along each axis. The legends show the IV-weighted average and standard deviation.}
    \label{fig:iridium_spectral_index_scatter_w_histogram}
\end{figure}

The measured fluxes of the Iridium satellites lie on the lower end of the model as shown in Figure \ref{fig:iridium_flux_vs_los_dist}, where the minimum area is given by a single antenna panel.
This remains true, even when assuming a polished aluminum emissivity of 0.03 indicating that the minimum cross-sectional area of a single antenna panel may be too large.
Given the unique panel configuration of Iridium satellites, exemplified in Figure 4 of \cite{maley03}, this may make sense if panels are reflecting into space rather than back to Earth.
The 90$^{\text{th}}$-percentile flux uncertainty for these observations is 0.1~Jy at 150~GHz.
There is a large scatter in the spectral indices, likely due to the low signal-to-noise of each observation, as shown in Figure \ref{fig:iridium_spectral_index_scatter_w_histogram}.
The mean and std. err. spectral indices are roughly consistent with thermal emission, namely, $\alpha^{95}_{150}$=1.99$\pm$0.03 and $\alpha^{150}_{220}$=2.09$\pm$0.04.

\subsection{Intentional Emission}
\begin{figure}[]
\centering
\includegraphics[width=1.\linewidth]{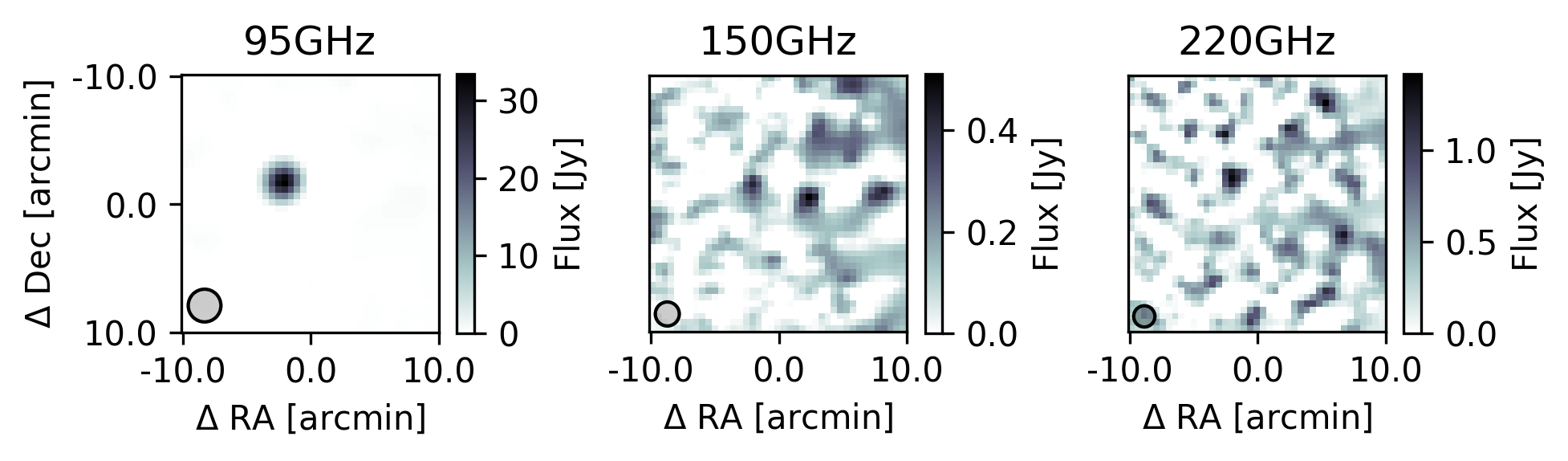}
\caption{Co-moving maps of the CloudSat satellite centered on the nominal, TLE-derived position of the satellite. The bright \SI{95}{\giga\hertz} emission is caused by the satellite's \SI{94.05}{\giga\hertz} cloud detecting radar. The SPT-3G beam is shown as a shaded 2$\sigma$ contour in the lower-left.}
\label{fig:cloudsat_maps}
\end{figure}

Another concern is intentional in-band emission, as with the \SI{94.05}{\giga\hertz} radar from the CloudSat instrument.
The emitted power is much higher than from typical astrophysical sources, so much so that it has been the discussion of some radio observatories using sensitive mixer electronics that can be damaged if observed in the main beam (ALMA Memo No. 504).
One SPT-3G observation of CloudSat occurred at a distance of 1264~km and is shown in Figure \ref{fig:cloudsat_maps}.
From Equation \eqref{eq:cloudsat_power}, the expected \SI{95}{\giga\hertz} power received by SPT-3G at this distance is 2.5~pW, which is 120~Jy in flux density units.

The SPT-3G measurement is 32~Jy in the \SI{95}{\giga\hertz} band but only 0.4 and 1.2~Jy at 150 and \SI{220}{\giga\hertz}, respectively.
The latter two fluxes are consistent with thermal blackbody radiation from part of the $\sim 16$~m$^2$ surface area.
The measured flux confirms that, in agreement with the back-of-the-envelope calculation from Section \ref{sec:emission_mechanisms}, even the far sidelobes of such a radar can be extremely bright.

\subsection{TLE Precision}
\label{sec:tle_precision}
Although many of the examples given above are centered on the brightest pixel for illustrative purposes, when constructing satellite-centered maps, the satellites appear up to several arcminutes offset even when using the nearest-in-time TLE epoch to construct the ephemerides.
No correlation was found between the amplitude or direction of the offset in the map and how far in time from the TLE epoch the satellite was observed.
Similar offsets have been observed by others (see for example the discussion on the measured BlueWalker TLE decay in \citealt{nandakumar2023}) and may be a consequence of low TLE precision or unmodeled perturbations such as orbital maneuvers.
This may be more important for LEO satellites where atmospheric drag is larger. 

Examples of the positional offsets observed by SPT-3G are presented in co-moving \SI{150}{\giga\hertz} maps in Figures \ref{fig:starlink_gridmap}-\ref{fig:lvm3_gridmap} in Appendix \ref{app:tle_precision}.
Only observations that have signal-to-noise greater than 5 are shown, though for Starlink observations it is clear that there is localized noise contamination at the level of the 150~GHz Starlink flux.
More work in reducing the position uncertainty remains an important next step for a thorough analysis of all satellites and is a priority of the International Astronomical Union's Center for the Protection of the Dark and Quiet Sky from Satellite Constellation Interference (CPS, \citealt{cps2024}).
If the positional uncertainties can be reduced to a few arcseconds, CMB observatories could stack maps of dim satellite constellations (such as OneWeb) in order to more accurately measure their millimeter properties.
Furthermore, optical observatories will likely require high precision position measurements to perform accurate satellite streak avoidance and removal.
CPS guidelines from 2021 requested publicly-available, live-updated telemetry information for LEO satellites with accuracy of 6 arcminutes, with a desired goal to increase the accuracy by a factor of 10 \citep{cps2021}.
Steps to improve accuracy require a combination of observatory measurements and metadata provided by the satellite operators as suggested in the SATCON2 report (see \citealt{satcon2}).

\section{The Effects of Satellites on CMB Survey Science}
\label{sec:effects_on_cmb_surveys}
The CPS has been actively working toward understanding and mitigating satellite constellation interference, primarily at optical and radio wavelengths.
This work expands upon that effort to seek to understand whether current and future mega-constellations will cause a noticeable effect on CMB survey science.
There are two likely scenarios to consider.
The first is that bright satellites (above 2~Jy at \SI{150}{\giga\hertz}) will be flagged by the TOD glitch finding step.
After glitch flagging, the satellite would not directly inject noise or artificial signals, but would cause some fraction of bolometers to be dropped from the scan thus causing an effective loss in observing efficiency.
The second scenario is that sub-glitch-threshold satellites can increase the noise in the maps and may do so in non-uniform ways.

To understand the scale of this problem, the worst-case limit is taken in which the glitch-removal is not active and a satellite is extremely bright (i.e. like the LVM3 upper stage described in Section \ref{sec:lvm3_thermal_emission}).
Assuming a flux of 10~Jy at \SI{150}{\giga\hertz} (a reasonable selection from Figure \ref{fig:gslv_flux_vs_distance}), this would inject a streak in the single-scan maps with an amplitude reduced by the number of detectors which observe the same sky position but not the satellite.
In general, a static point on the sky traverses between 20-50 focal plane pixels during a scan.
Each pixel contains 6 detectors, two in each observing band, for an average of 70 detectors observing a stationary sky location per scan per band.
This means that the fast-moving 10~Jy satellite will be averaged by a factor of 35, to 280~mJy in the single-scan map.
That level of signal is still quite large, however as of the writing, there are no real-time analyses performed on single-scan maps.

Perhaps a more useful quantity is the amplitude of this satellite in the single-observation map.
Due to the raster scan strategy, a single location on the sky is generally observed in 20-30 scans during a single observation.
Dividing by an average of 25 scans, the satellite would make an 11~mJy streak in the map.
The length of this streak would be of order 1~degree over a beamwidth of approximately 1~arcminute for SPT-3G.
The noise injected into this streak, albeit over a small region of the sky, is comparable to typical beam-scale map noise which is roughly 10~mJy at \SI{150}{\giga\hertz}.

With the standard glitch flagging, signals above 2~Jy are removed, thus a more realistic maximum noise injection scenario is a 2~Jy residual resulting in a single-observation averaged signal of just 2~mJy.
When added in quadrature to the map noise, this represents an increase of $\sim$5\%, contained to only a few beams across the focal plane.

\begin{figure*}[h!]
\begin{minipage}[c]{0.7\linewidth}
\includegraphics[width=\linewidth]{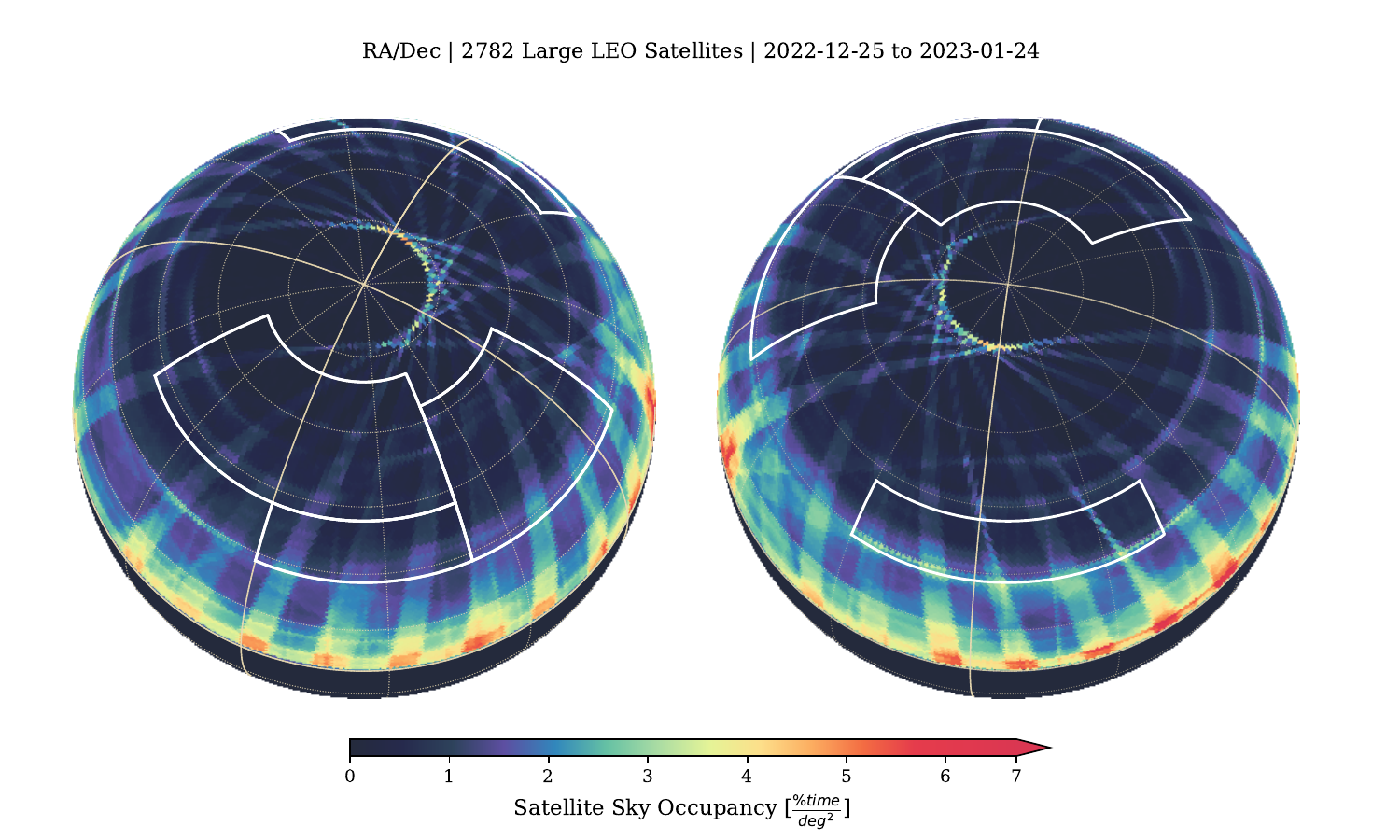}
\end{minipage}
\hfill\hspace{-.75in}
\begin{minipage}[c]{0.35\linewidth}
\includegraphics[width=\linewidth]{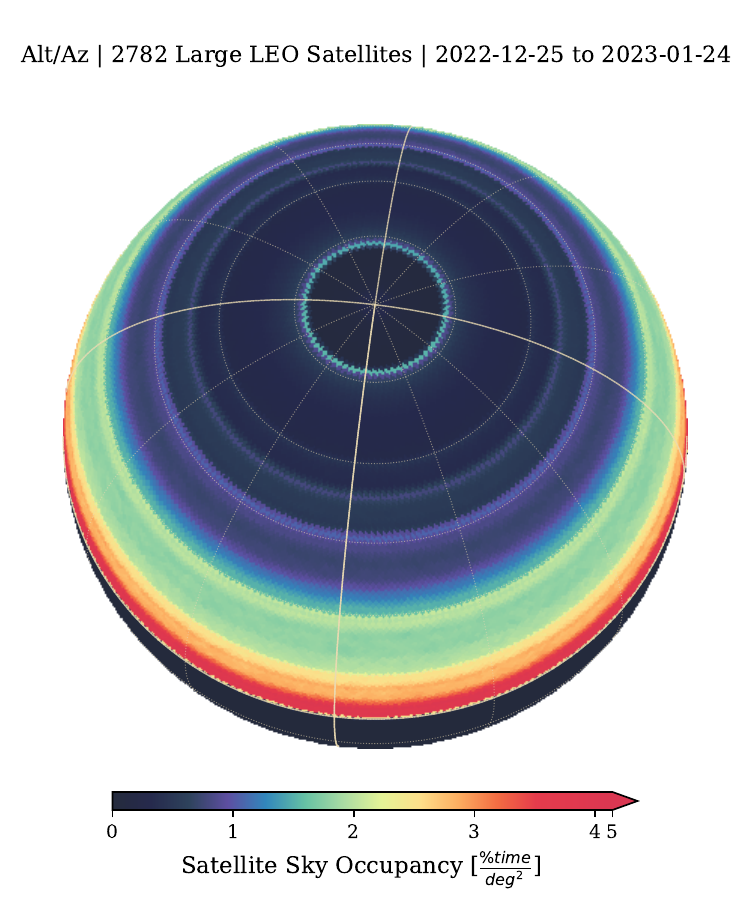}
\end{minipage}%
\caption{ Satellite fractional fill factor, defined in Section \ref{sec:effects_on_cmb_surveys}, averaged from December 25$^{\text{th}}$ 2022 through January 24$^{\text{th}}$ 2023. The entire southern sky is shown in ecliptic coordinates in the left two images with all observed SPT-3G survey regions outlined in white and shown in more detail in Figure \ref{fig:fractional_satellite_fill_fields}. Great circles are drawn in yellow at RA=0,180 and RA=90,270 degrees with thin dotted grid lines every 15 degrees. The right image shows the fractional fill factor in local coordinates (altitude vs. azimuth at the South Pole).}
\label{fig:fractional_satellite_fill_sky}
\end{figure*}

\begin{figure*}[h!]
\centering
\includegraphics[width=0.9\linewidth]{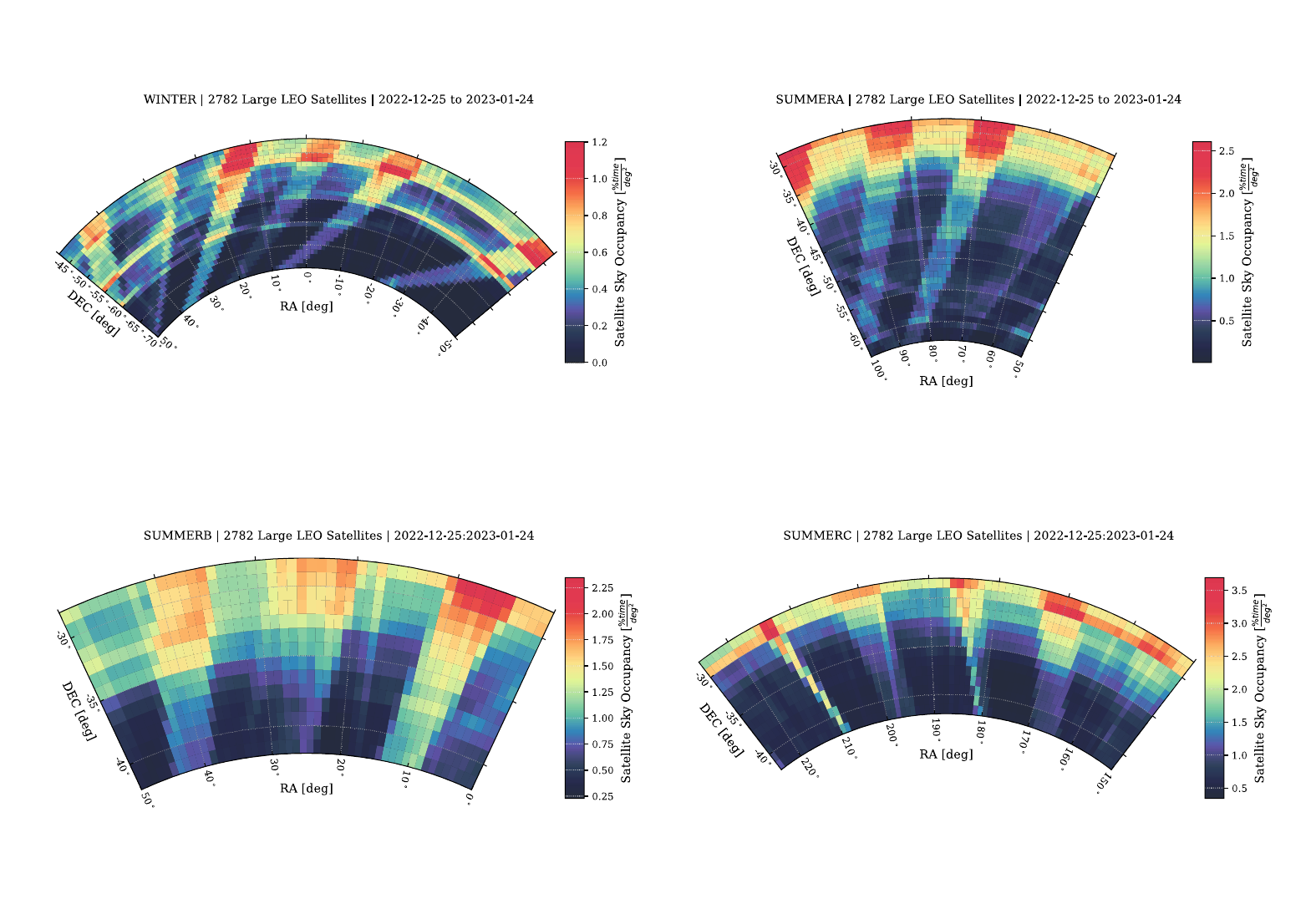}
\caption{Satellite fractional fill factor, defined in Section \ref{sec:effects_on_cmb_surveys}, calculated for the SPT-3G 1500d survey region (labeled here as WINTER) and three austral summer fields and averaged from December 25$^{\text{th}}$ 2022 through January 24$^{\text{th}}$ 2023. Only ``Large" RCS, LEO satellites are included.}
\label{fig:fractional_satellite_fill_fields}
\end{figure*}

Figure \ref{fig:fractional_satellite_fill_sky} shows the fractional fill factor, which represents the fraction of time, per second, that a square degree of sky contains a ``Large" LEO satellite, averaged over the course of a month. 
The calculation is based on satellite ephemerides sampled at a 1 second cadence and mapped onto a nside=64 ($\sim$0.84 square degree resolution) HEALPix grid. 
For each sample, if a satellite is found within a given pixel, that pixel's count is incremented by one. 
The total counts for each pixel are then divided by the total number of samples in the month to obtain the fractional fill factor. 
Above the horizon at the South Pole, some regions of the sky exhibit a fractional fill factor of order 10\% deg$^{-2}$ s$^{-1}$, with the fill factor increasing toward the horizon.
Within the SPT-3G survey regions, that number peaks around a few percent.
Figure \ref{fig:fractional_satellite_fill_fields} shows a zoom-in on the SPT-3G observing fields, highlighting the non-uniformity of satellite distribution in RA, Dec.
Since the primary cosmological survey science relies on averaging down noise over many years, the impact from satellites becomes less-pronounced, similar to the averaging of atmospheric noise.

Assuming an extreme scenario where these satellites are all above 2~Jy, the expected signal contribution to a single-observation map is 2~mJy per beam through which the satellite traverses.
Because a satellite traverses on average 40 focal plane pixels during a pass, in each $\sim$\SI{2}{\hour} observation there are a total of $\mathcal{N} \times$40 beams affected in the final 400~deg$^2$ subfield map (per band), where $\mathcal{N}$ is the number of satellites which pass through the field-of-view during that observation.
This covers a map fraction
\begin{equation}
    f=\mathcal{N} \left ( \dfrac{40~\text{arcmin}^2}{ 400~\text{deg}^2} \right ) \left (\dfrac{1~\text{deg}^2}{3600~ \text{arcmin}^2}\right ).
    \label{eq:map_frac_satstreaks}
\end{equation}
Using the highest-density region from Figure \ref{fig:fractional_satellite_fill_sky} with a fill factor of order 10\% per square degree per second (a pessimistic estimate), the number of observed satellite passes per observation can be written as
\begin{equation}
    \mathcal{N} = 2~\text{deg}^2 ~\left (\dfrac{0.1}{\text{deg}^2~\text{s}} \right ) \left ( 2~\text{hr}~\dfrac{\SI{3600}{\second}}{1\text{hr}} \right ) =1440,
\end{equation}
where the 2~deg$^2$ factor is the area of the SPT-3G focal plane.
Plugging this into Equation \eqref{eq:map_frac_satstreaks}, the fraction of the single observation map with an extra 2~mJy of satellite-induced signal is $f=0.025$.

The SPT-3G RMS map noise in single observation maps is approximately 10~mJy per beam at \SI{150}{\giga\hertz}, thus the average map noise can be written as
\begin{equation}
    \bar{N} = \sqrt{(0.975)10^2 + (0.025)12^2}~\text{mJy} 
    \approx 10.05~\text{mJy}.
\end{equation}
This is only a 0.5\% increase in total map noise in the most pessimistic case.
Combined with the fact that these over-dense regions move over time and the mean increase in noise will be less than this extreme localized case, this is not a significant noise increase.

This estimate is also a limit in which all ``Large" RCS LEO satellites are above the 2~Jy threshold and are present in one of the survey regions, which is not realistic; thus the true increase in map noise is expected to be much less.

Recent work in short-duration time-domain searches for astrophysical transients using SPT-3G has brought up the question of whether satellites may also pose as a source of false-positive transient signals.
While known satellites can be masked from the data, any non-public satellite or debris with poor orbital constraints could cause a bright (and brief) specular reflection.
If the duration of the specular flare were short enough (of order a few milliseconds) it might appear to be consistent with a stationary source.
While this type of a flare would be unlikely, short duration transients such as gamma-ray burst reverse shocks are also extremely rare (none have been detected in 4~years of SPT-3G 1500d observations Guns et al., in preparation).

In the future, CMB surveys such as SO and CMB-S4 \citep{abazajian19} will be subject to mega-constellations of satellites.
The proposed 48,000 member Starlink constellation, once complete, will contain roughly 1 satellite per square degree on the sky.
The South Pole, however, may remain a relatively clean site since only LEO orbits with very nearly polar inclinations are visible in the survey region.

Work remains to understand the positional offsets between expected satellite positions derived from the TLEs and their recovered locations, which can be discrepant by many arcminutes (see Section \ref{sec:tle_precision} and Appendix \ref{app:tle_precision}).
A thorough search for non-thermal emission has not yet been conducted and may reveal a subset of satellites which are actively emitting in-band.
Similarly, only total intensity has been measured in this work.

\section{Conclusions}
\label{sec:conclusions}
This work represents the first systematic analysis of millimeter thermal emission of Earth-orbiting satellites using a ground-based CMB survey instrument.
Although the measured thermal emission places large, nearby satellites amongst the brightest millimeter sources observed by CMB survey instruments, they do not currently lead to a significant increase in the survey map noise.
No observations of bright flares due to specular reflection of the Sun were detected over several hundred observed satellite passes.
Actively transmitting satellites can be extremely bright, even in the far-field sidelobes, which means that simple transmission exclusion zones may not be enough to prevent interference.
Future efforts should be undertaken to understand which satellites are producing non-thermal in-band emission.

Satellite ephemerides constructed using the nearest-epoch TLEs are shown to have offsets as large as several arcminutes.
This limitation will make satellite prediction, masking, or cross-matching difficult for upcoming observatories, particularly as the number of LEO satellites continues to grow.

Significant work will be required to break the degeneracy between temperature, geometric cross-section, and emissivity for these objects.
For example, one may attempt to measure the thermal emission above the Rayleigh-Jeans tail in order to measure or place limits on the temperature or to perform high-resolution imaging to measure the geometric cross-section.
The latter will require a much more accurate method for calculating satellite ephemeris than the current schema using TLEs.

An analysis of the possible effects of current satellites on CMB survey data showed that satellite thermal emission is not a significant component of survey map noise, even in the scenario where every satellite is above the glitch threshold.
If large LEO constellations increase the number of on-orbit satellites by an order of magnitude, as expected in the coming decade, this increase in survey map noise may become non-negligible and reach the few percent level.
Researchers conducting short-duration transient searches should be aware of the presence of satellites and will need to cross-check candidate events with the caveat that a large (several arcminutes) matching radius would be needed.
In general, for map-based transient searches, these fast-moving satellites get averaged down in fixed-sky coordinates to negligible levels over the $\sim$hours-long observation, however future searches directly in the time-ordered data may require more careful analysis.

\section{Acknowledgments}
\label{sec:acks}
The South Pole Telescope program is supported by the National Science Foundation (NSF) through awards OPP-1852617 and OPP-2332483.
Partial support is also provided by the Kavli Institute of Cosmological Physics at the University of Chicago.
This research made use of resources provided by the Open Science Grid \citep{pordes07, sfiligoi09}, which is supported by the NSF award 1148698, and the U.S. Department of Energy's Office of Science.
The data analysis pipeline also uses the scientific python stack \citep{hunter07, jones01, vanDerWalt11}.
The satellite orbital parameters and TLEs were downloaded from the space-track.org webserver provided by the USAF 18th Space Defense Squadron.
Argonne National Laboratory’s work was supported by the U.S. Department of Energy, Office of High Energy Physics, under contract DE-AC02-06CH11357.  
Work at Fermi National Accelerator Laboratory, a DOE-OS, HEP User Facility managed by the Fermi Research Alliance, LLC, was supported under Contract No. DE-AC02-07CH11359.

\bibliographystyle{aasjournal.bst}
\bibliography{satellites}

\clearpage

\appendix
\section{Satellite Elevation Calculation}
\label{app:satellite_elevation_calc}

\begin{figure}[h]
\centering
\includegraphics[width=.95\linewidth]{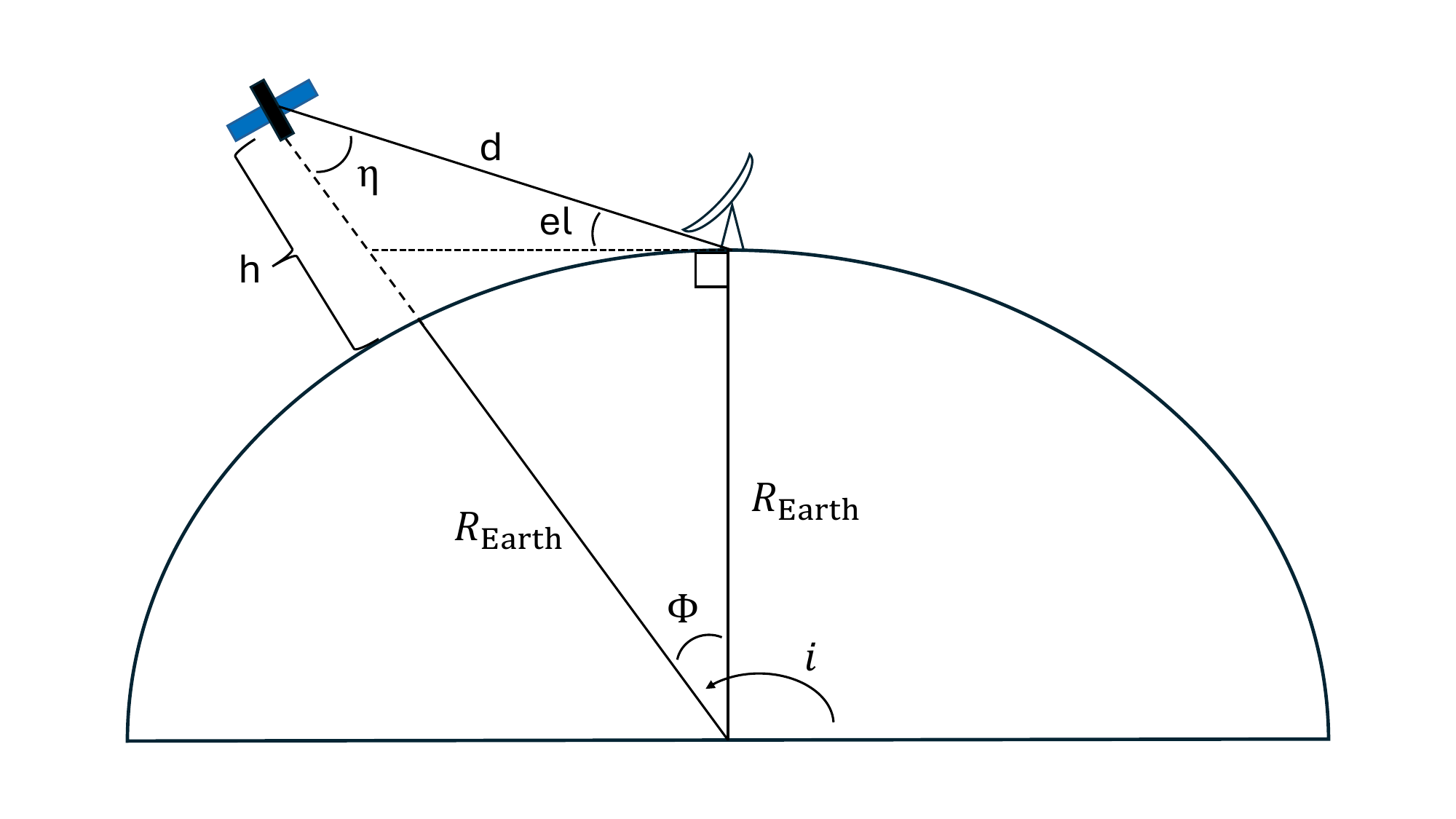}
\caption{Cartoon diagram showing the geometry of a satellite as observed from the South Pole. The satellite has orbital inclination, $i$, and altitude, h. The line-of-sight distance to the SPT is d. This simple picture can be used to calculate if a satellite with given orbital inclination and apogee will rise above the SPT-3G survey horizon.}
\label{fig:sat_el_calc}
\end{figure}

Due to the special geographic location of the South Pole, there are many satellites which will never rise above the horizon.
Figure \ref{fig:sat_el_calc} shows a cartoon graphic depicting the geometry of a satellite as observed from the South Pole.
The radius of Earth, R$_{\text{Earth}}$, is taken to be \SI{6358}{\kilo\meter} and the elevation of the SPT above sea level (\SI{2.8}{\kilo\meter}) is ignored.

The triangle in Figure \ref{fig:sat_el_calc}, defined by sides (h+R$_{\text{Earth}}$, R$_{\text{Earth}}$, d) with internal angles ($\eta$, $\Phi$, el+\SI{90}{\degree}), can be used to solve for the elevation.
Using the law of sines,
\begin{equation}
    \dfrac{\text{sin}(\SI{90}{\degree}+\text{el})}{\text{h+R}_{\text{Earth}}} =  \dfrac{\text{sin}(\eta)}{\text{R}_{\text{Earth}}}, \label{eq:lawofsines}
\end{equation}
and the fact that the internal angles add up to 180 degrees, $\eta$ can be written in terms of elevation and inclination, $i$.
Since $\Phi=i-$\SI{90}{\degree}, Equation \eqref{eq:lawofsines} becomes

\begin{equation}
    \dfrac{\text{sin}(\SI{90}{\degree}+\text{el})}{\text{h+R}_{\text{Earth}}} =  \dfrac{\text{sin}(\SI{180}{\degree}- ( i-\SI{90}{\degree}) - (\text{el}+\SI{90}{\degree}))}{\text{R}_{\text{Earth}}},
\end{equation}
which can be rearranged using some trig identities to obtain
\begin{equation}
    \dfrac{\text{cos}(\text{el})}{\text{h+R}_{\text{Earth}}} =  \dfrac{\text{sin}(i)\text{cos(el)}+\text{cos}(i)\text{sin(el)}}{\text{R}_{\text{Earth}}}.
\end{equation}
Solving for elevation gives
\begin{equation}
    \text{el} = \text{tan}^{-1} \left[ \dfrac{\text{cos}(i)~\text{R}_{\text{Earth}}}{\text{h+R}_{\text{Earth}}}- \text{tan}(i) \right ]. \label{eq:sat_elevation}
\end{equation}
Figure \ref{fig:sat_visibility} shows the calculated elevation from Equation \eqref{eq:sat_elevation} of known satellites which are possibly visible above the horizon from the South Pole using apogee as an upper limit for altitude.

\section{Thermal (Blackbody) Emission Calculation}
\label{app:thermal_emission_calc}

\setcounter{table}{0}
\renewcommand{\thetable}{\Alph{section}\arabic{table}}
\renewcommand*{\theHtable}{\thetable}

The total emission from a blackbody at temperature $T$ can be calculated using the Stefan-Boltzmann (SB) law
\begin{equation}
P = A \sigma_{\text{SB}}~T^4,    \label{eq:sb_law} 
\end{equation}
where $A$ is the surface area and $\sigma_{\text{SB}}=5.67\times10^{-8}$~Wm$^{-2}$K$^{-4}$ is the Stefan-Boltzmann constant.
The in-band emission is calculated by integrating the spectral radiance over the instrument bandpass.
The spectral radiance can be calculated using the Planck blackbody function, defined as
\begin{equation}
    \mathcal{B}(\nu,T) = \dfrac{2h\nu^3}{c^2}\dfrac{1}{\text{exp}\left ( \dfrac{h\nu}{k_BT} \right )-1}, \label{eq:planck_func}
\end{equation}
where $h$, $c$, and $k_B$ are the Planck constant, speed of light, and Boltzmann constant, respectively.

The fraction of the total emission which is radiated in-band is then calculated by dividing by \eqref{eq:sb_law}, namely,
\begin{equation}
    f_b^T = \dfrac{\pi \int\mathcal{B}(\nu,T)d\nu }{\sigma_{\text{SB}} T^4}, \label{eq:frac_em}
\end{equation}
where $f_b^T$ is the fractional emittance in band $b$ at temperature $T$ and the integral is over the bandwidth.
The factor of $\pi$ in the numerator reflects the fact that the SB constant is calculated by integrating Equation \ref{eq:planck_func} over the hemisphere.
For the Sun, using a blackbody temperature of 5780~K, the total emittance is $6.33\times10^7$~W/m$^2$.
Integrating over the surface area of the Sun, whose radius is $6.96\times10^8$~m, and radiating isotropically to a distance of 1~AU (1.5$\times10^{11}~$m), the received power per unit area at Earth is 1370~W/m$^2$.
Similarly, for a satellite in thermal-equilibrium with Earth (assumed to be a 300~K blackbody), the total emission (from Equation \ref{eq:sb_law}) is 460~W/m$^2$. 
The fractional emittance in each SPT-3G band at these two temperatures is given in Table \ref{tab:fractional_emittance_per_band} and assumes the top-hat bandpasses given in Section \ref{sec:telcam}.

\begin{table}[ht]
\caption{Fractional emittance ($f_b^T$, Equation \ref{eq:frac_em}) in each SPT-3G band for a blackbody of temperature 5780~K and 300~K.}
\centering
\begin{tabular}{l c c c}
\toprule
 & \textbf{\SI{95}{\giga\hertz}}& \textbf{\SI{150}{\giga\hertz}} & \textbf{\SI{220}{\giga\hertz}} \\
\midrule
300~K & 1.5~x~10$^{-7}$ & 4.4~x~10$^{-7}$ & 1.6~x~10$^{-6}$ \\
5780~K & 2.1~x~10$^{-11}$ & 6.2~x~10$^{-11}$ & 2.3~x~10$^{-10}$ \\
\bottomrule
\end{tabular}
\label{tab:fractional_emittance_per_band}
\end{table}

The expected in-band emittance from a 300~K blackbody and from the Sun, incident on an object at 1~AU, can now be calculated and is given in Table \ref{tab:inband_emittance} in units of W/m$^2$ for each of the SPT-3G frequency bands.
Using the fact that at these frequencies and for the expected temperatures experienced by Earth-orbiting satellites, the Rayleigh-Jeans limit is valid and the emitted power is simply proportional to temperature as
\begin{equation}
    \epsilon_b^T \equiv f_b^T \cdot \sigma_{\text{SB}} T^4 = \epsilon_b^{300K} \left( \dfrac{T}{300~\text{K}} \right), \label{eq:emittance}
\end{equation}
where $\epsilon_b^T$ is the emittance in band $b$ for a blackbody of temperature $T$.

\begin{table}[h!]
\caption{In-band emittance ($\epsilon_b^T$) from a surface for each SPT-3G band from a blackbody at 300~K and reflected sunlight at 1~AU. All values are in units of W/m$^2$.}
\label{tab:inband_emittance}
\centering
\begin{tabular}{l c c c}
\toprule
& \textbf{\SI{95}{\giga\hertz}}& \textbf{\SI{150}{\giga\hertz}} & \textbf{\SI{220}{\giga\hertz}} \\
\midrule
300~K & 6.7~x~10$^{-5}$ & 2.0~x~10$^{-4}$ & 7.5~x~10$^{-4}$ \\
Sun (1AU) & 2.8~x~10$^{-8}$ & 9.0~x~10$^{-8}$ & 3.1~x~10$^{-7}$ \\
\bottomrule
\end{tabular}
\end{table}

It is now straight-forward to calculate the received power at the SPT primary aperture, $\pi r^2$, due to intrinsic thermal emission from a satellite with apparent cross-sectional area, $A$,  emissivity, $\varepsilon$, and temperature, $T$~(in K), at a distance, $d$:
\begin{equation}
P^{\text{therm}}_b= A ~
     \varepsilon_b~ 
     (1-\rho_b)~
     \epsilon_b^{300K}~ \left( \dfrac{T}{300~\text{K}} \right)  
     \left ( \dfrac{r}{d} \right)^2. \label{eq:300Kpower_intrinsic}
\end{equation}
$P^{\text{therm}}_b$ is the received thermal power from the satellite in band $b$ and $\rho_b$ is the band-dependent atmospheric opacity.
It is possible that the emissivity $\varepsilon_b$ is band-dependent, hence the subscript.
$\epsilon_b^{300K}$ is the in-band emittance at 300~K from Table \ref{tab:inband_emittance}, carrying the units W/m$^2$.
The final term is the effective fractional solid angle of the telescope aperture at $d$, relative to the hemisphere (i.e. the factor of $\pi$ in \ref{eq:frac_em}).

Since satellites are generally not perfect blackbodies and often have reflective thermal shielding, the Earth's own thermal emission may be reflected off of the satellite back to an observer on the ground.
The additional emission can be added to the thermal budget given the effective geometric form factor of the Earth, $F_\text{E}$, as seen from the satellite to give
\begin{equation}
P^{\text{therm}}_b= A ~
     \varepsilon_b ~
     (1-\rho_b)~
     \epsilon_b^{300K}~ \left( \dfrac{T}{300~\text{K}} \right)  
     \left ( \dfrac{r}{d} \right)^2~
     \left ( 1 + \dfrac{1-\varepsilon_b}{\varepsilon_b} ~F_{\text{E}}\right ).\label{eq:300Kpower}
\end{equation}
For reference, the geometric form factor of the Earth for an object with an altitude, $h$, is 
\begin{equation}
F_{\text{E}} = 1-\sqrt{\dfrac{h(h+2R_{\text{Earth}})}{R_{\text{Earth}}+h}}
\end{equation}
which varies from approximately 0.7 to 0.35 over the LEO altitude range (see \citealt{stevenson61}, pages 60-61).
Although the millimeter emissivity of the satellites is not known, it is clear from Equation \ref{eq:300Kpower} that the thermal emission will be maximized as $\varepsilon_b$ approaches 1, and reduced to $F_{\text{E}}$ times the nominal 300~K blackbody emission as $\varepsilon_b$ approaches 0.

The received power can be converted to flux density via 
\begin{equation}
    S^{\text{therm}}_{b} =\dfrac{P^{\text{therm}}_b}{\Delta\nu_b ~ \pi r^2},
\end{equation}
where $\Delta \nu_b$ is the bandwidth of band $b$. 
Plugging in Equation \eqref{eq:300Kpower}, the telescope apertures cancel, leaving
\begin{equation}
    S_b^{\text{therm}} =  \dfrac{A ~ \varepsilon_b \ (1-\rho_b) ~ \epsilon_b^{300K} \left( \dfrac{T}{300~\text{K}} \right) \left ( 1 + \dfrac{1-\varepsilon_b}{\varepsilon_b} ~F_{\text{E}}\right )}
     {\Delta\nu_b ~\pi d^2 }. \label{eq:sat_thermal_jansky_general}
\end{equation}

Using Table \ref{tab:inband_emittance}, the incident power in the SPT-3G \SI{150}{\giga\hertz} band can be calculated for a general satellite.
For SPT-3G, the standard calibration pipeline also corrects for atmospheric opacity during the time of observation by measuring the flux of a known, non-varying astrophysical source. 
Equation \eqref{eq:sat_thermal_jansky_general} is therefore simplified by setting $\rho_b=0$.
An assertion is made that $T=300$~K and the degeneracy with $A$ and $\varepsilon_b$ is absorbed into an effective cross-section.
This represents the cross-sectional area that an object would have if it were a perfect 300~K blackbody and there was no atmospheric extinction.
Defining the effective cross-section as $\mathcal{A}$, 
\begin{equation}
    P^{\text{therm}}_{\text{150GHz}}(\mathcal{A},d) = 5~\text{fW} \left(\dfrac{\mathcal{A}}{1~\text{m}^2}\right) \left(\dfrac{1000~\text{km}}{d}\right)^2.\label{eq:150GHz_thermal_received_power}
\end{equation}
Similarly, the flux density measured at \SI{150}{\giga\hertz} can be estimated as
\begin{equation}
   S^{\text{therm}}_{\text{150GHz}}(\mathcal{A},d) = 200~\text{mJy} \left(\dfrac{\mathcal{A}}{1~\text{m}^2}\right) \left(\dfrac{1000~\text{km}}{d}\right)^2. \label{eq:150GHz_thermal_flux_density}
\end{equation}

\vspace{10pt}
\section{Specular Reflection Calculation}
\label{app:specular_flare_calc}
Before calculating the emittance due to a specular reflection, it is important to calculate the probability of actually observing one.
Because the angular diameter of the Sun is 0.5~deg, a specular reflection off of a large surface would radiate into a cone of radius $\theta_{\text{Sun}}$=0.25~deg.
If one decreases the size of the reflector, eventually diffraction will become important as well.
Including diffraction in quadrature, the reflection subtends an angle
\begin{equation}
   \theta = \sqrt{\left ( 1.22\dfrac{\lambda}{D} \right ) ^2+(\theta_{\text{Sun}})^2}, \label{eq:ref_angle}
\end{equation}
where $\lambda$ is the observed wavelength and $D$ is the effective size (diameter) of the reflector.
The radius of the diffraction beam at Earth is $d\theta$ in the small angle approximation, where $d$ is the line-of-sight distance to the satellite.
For a 2~mm wavelength, a 1~m effective diameter, and a LOS distance of 1000~km, the spot radius at Earth is 5~km and is dominated by $\theta_{\text{Sun}}$ for a reflector with $D>\sim$0.6~m.

The probability that this diffraction disk is observable at a location on Earth is calculated assuming the orientation of the reflector is random.
The area illuminated by the reflected light is 
\begin{equation}
    \Omega = 2\pi\left(1 - \cos\theta\right)~\text{sr} = 4\pi\sin^2 \left(\frac{\theta}{2}\right)~\text{sr},
\end{equation}
and therefore the fraction of the total sky-area illuminated is 
\begin{equation}
    f_{\text{sky}}=\dfrac{\Omega}{4\pi}=  \sin^2\left(\frac{\theta}{2}\right).
\end{equation}
In the limit that the illumination angle is small, 
\begin{equation}
    f_{\text{sky}}\approx \left(\frac{\theta}{2}\right)^2 = \frac{1}{4}\left( \left(1.22\dfrac{\lambda}{D}\right)^2+(\theta_{\text{Sun}})^2\right).
\end{equation}
Thus for a 1~m diameter reflector and \SI{150}{\giga\hertz} light ($\lambda=2$~mm), the fraction of sky illuminated by the specular reflection is 5$\times10^{-6}$.
Of course, this is the probability of being able to observe a specular reflection for a single, randomly oriented surface above your horizon at any given time.

Satellites are generally not monolithic panels and may contain many smaller sub-structures. 
If each satellite consists of N panels with area A/N, the number of observed reflections will scale with N.
The smaller panel sizes mean a larger diffracted beam area, thus the probability of observing a reflection increases with another factor of N. 
The amplitude, however, scales with N$^{-1}$ because of the smaller panel size and with N$^{-1}$ due to the larger diffracted beam area. 
In this schema, the probability of observing specular reflections is proportional to N$^2$ with an amplitude proportional to N$^{-2}$.

This instantaneous probability can be converted to a rate by considering the number of satellites which pass above the SPT horizon, defined in this work as \SI{28}{\degree} elevation.
As simplifying assumptions, each satellite's orbital period is assumed to be 100~minutes, each satellite can only be observed once per orbit, and orbits are randomly oriented on the sky.
The final assumption turns out not to be applicable to several sun-synchronous orbits as shown in Section \ref{sec:effects_on_cmb_surveys}, which may increase or decrease the probability of observation depending on whether they move through the survey region during each orbit or not.
The sky area above the SPT horizon is approximately 18000~deg$^2$ and the SPT-3G focal plane subtends $\sim$2~deg$^2$.
Thus from Section \ref{sec:satellites} there are 2,286 satellites above the SPT-3G horizon, each possibly observed once per 100 minutes with an observing efficiency of 2~deg$^2$/ 18000~deg$^2$
This results in 0.15 satellites per hour observed by SPT-3G as a lower limit.
For an upper limit, each satellite is assumed to pass through one of the SPT-3G survey regions on each orbit, changing the effective sky area to 4500~deg$^2$ and yielding a satellite observation rate of 0.6 satellites per hour.
If each satellite had a single reflective surface, the expected rate is of order 10$^{-6}$ per hour which is about 1 observed specular reflection in 100 years!

Although the random orientation assumption may not be appropriate for operational satellites, one can extrapolate from the well-known example of specular reflections off of Iridium satellite antennae, colloquially known as ``Iridium flares".
Following \cite{hainaut20}, the optical flare rate of the 66 satellite Iridium constellation was a few per night and each satellite had three large reflecting panels.
Assuming each of the satellites visible to SPT-3G has only one such panel and with an observed rate scaled from Iridium (one-third of $\sim$three flares per 66 satellites per night), the expected rate of satellite flares is about 35 over the observable sky per night.
This means that the expected number per SPT-3G field-of-view is 0.004 observed flares per night, which is roughly 1 observed flare per year.

Being so rare, one may ask if it is even important to consider the impact of specular reflection flares.
The emittance calculated from diffuse solar reflection is only 0.1\% that of thermal radiation in the SPT-3G bands (Table \ref{tab:inband_emittance}), however if the reflection is specular, a much higher fraction of the total incident radiation may be beamed into the receiver.
This would cause a brief and bright signal which may masquerade as a short-duration astrophysical transient.
Using the same procedure as for thermal emission in Appendix \ref{app:thermal_emission_calc}, 
\begin{equation}
P_b^{\text{spec}} = A ~ (1-\varepsilon_b) ~ (1-\rho_b) ~ \epsilon_b^{\text{Sun}}\left ( \dfrac{r}{d ~ \theta} \right)^2 , \label{eq:Psun_ref}
\end{equation}
and
\begin{equation}
    S_b^{\text{spec}} = ~\dfrac{\epsilon_b^{\text{Sun}}}{ \pi \Delta\nu_b } ~ A ~ (1-\varepsilon_b) ~ (1-\rho_b)~ d^{-2} ~ \theta^{-2}, \label{eq:Ssun_ref}
\end{equation}
where $\theta$ is the subtended angle of reflection from Equation \eqref{eq:ref_angle} and is assumed to be within the small-angle regime.
Defining the reflecting surface area with the simple approximation that $A=\pi(D/2)^2$ and plugging Equation \eqref{eq:ref_angle} into \eqref{eq:Psun_ref}  and \eqref{eq:Ssun_ref}, the received power from specular reflection of the Sun becomes
\begin{equation}
P_b^{\text{spec}} \simeq \epsilon_b^{\text{Sun}} \left ( \dfrac{r}{d} \right)^2 \dfrac{A^2}{\lambda^2 +   A\theta_{\text{Sun}}^2}. \label{eq:Psun_ref_2}
\end{equation}
Optical opacity, $\rho$, has been calibrated out per the SPT calibration procedure and the surface is taken to be a perfect reflector, i.e. $(1-\varepsilon_b=1)$.
Using Table \ref{tab:inband_emittance} to calculate the power received at \SI{150}{\giga\hertz} ($\lambda$=2~mm) gives
\begin{equation}
P_{\text{150GHz}}^{\text{spec}}(A,d) \simeq 86~\text{fW} ~\dfrac{\left ( \dfrac{1000~\text{km}}{d} \right)^2 
                         \left(\dfrac{A}{1~\text{m}^2}\right)}
                         { 0.2 \left(\dfrac{1~\text{m}^2}{A}\right)+0.8}, \label{eq:Psun_ref_gen}
\end{equation}
and similarly the flux density is 
\begin{equation}
\begin{aligned}
S_\text{150GHz}^{\text{spec}}(A,d) \simeq 3.3~\text{Jy} ~\dfrac{\left ( \dfrac{1000~\text{km}}{d} \right)^2 
                         \left(\dfrac{A}{1~\text{m}^2}\right)}
                         { 0.2\left(\dfrac{1~\text{m}^2}{A}\right) + 0.8} .
\end{aligned} \label{eq:Ssun_ref_gen}
\end{equation}

\section{TLE Precision}
\label{app:tle_precision}
This appendix simply provides the figures which show the TLE-derived satellite-centered maps.
A satellite whose orbit is accurately described by the TLE should be found at the center of these maps, indicating that in the majority of cases, the satellites are offset by several arcminutes.

\begin{figure}[h!]
    \centering
    \includegraphics[width=.6\linewidth]{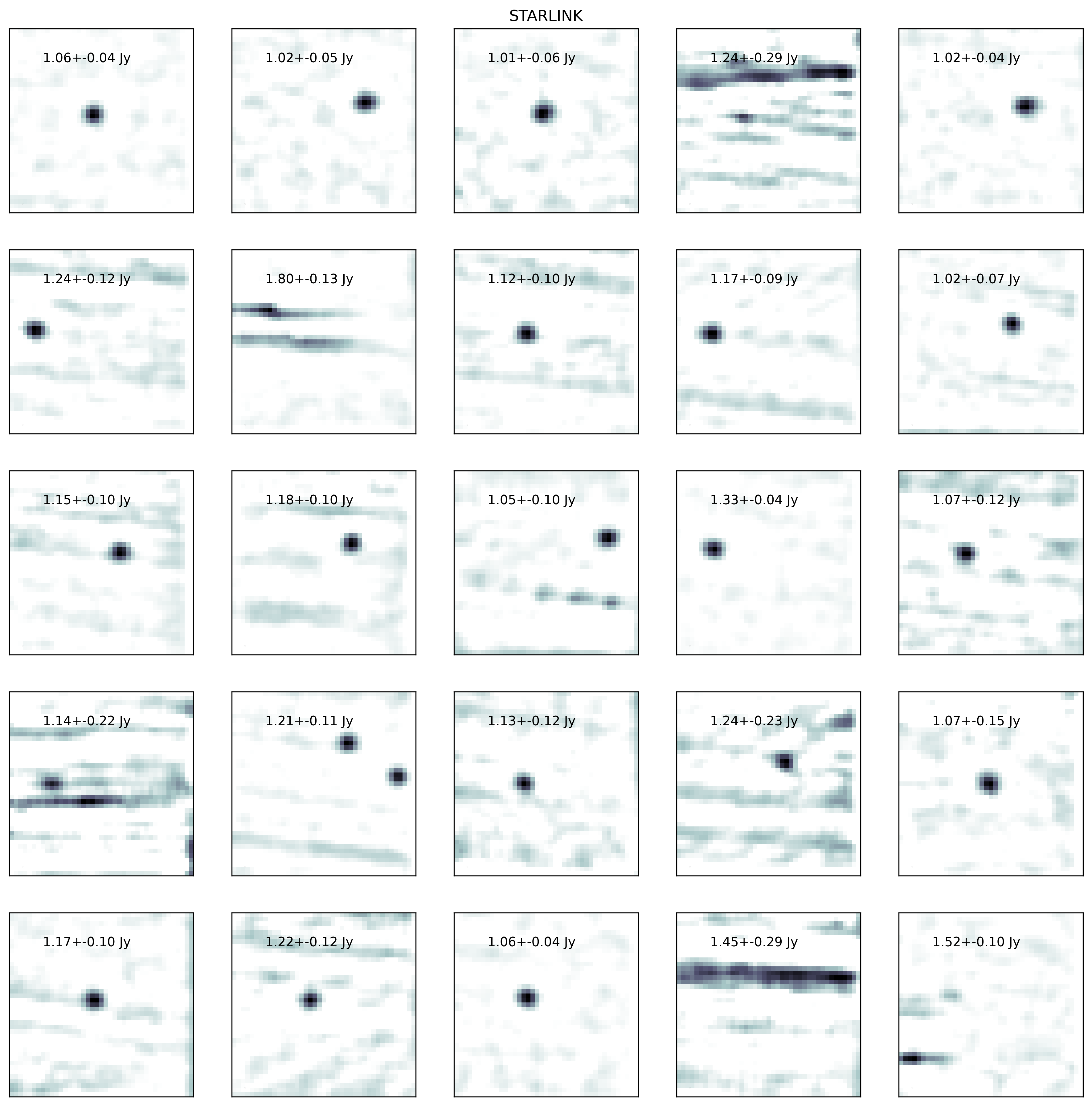}
    \caption{\SI{20}{\arcmin}$\times$\SI{20}{\arcmin} co-moving maps of 25 high signal-to-noise observations of Starlink satellites by SPT-3G at 150~GHz. Each map is centered on the nominal, TLE-derived position of the satellite. Some streaks are visible along the telescope scan direction and may be due to a nearby bright astrophysical source or localized readout noise.}
    \label{fig:starlink_gridmap}
\end{figure}

\begin{figure}[h!]
    \centering
    \includegraphics[width=0.75\linewidth]{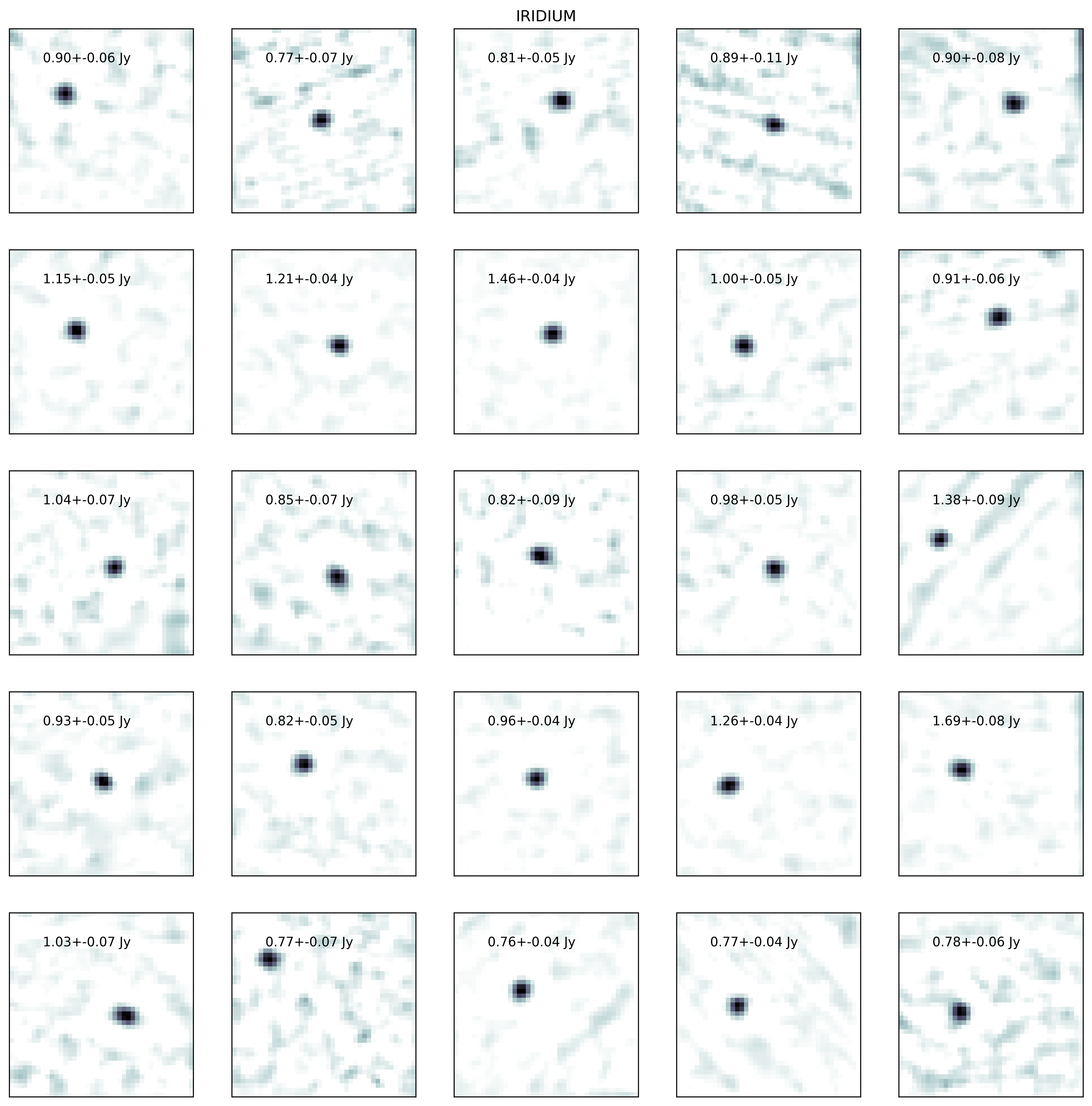}
    \caption{Same as Figure \ref{fig:starlink_gridmap} but for Iridium satellites.}
    \label{fig:iridium_gridmap}
\end{figure}

\begin{figure}[h!]
    \centering
    \includegraphics[width=0.75\linewidth]{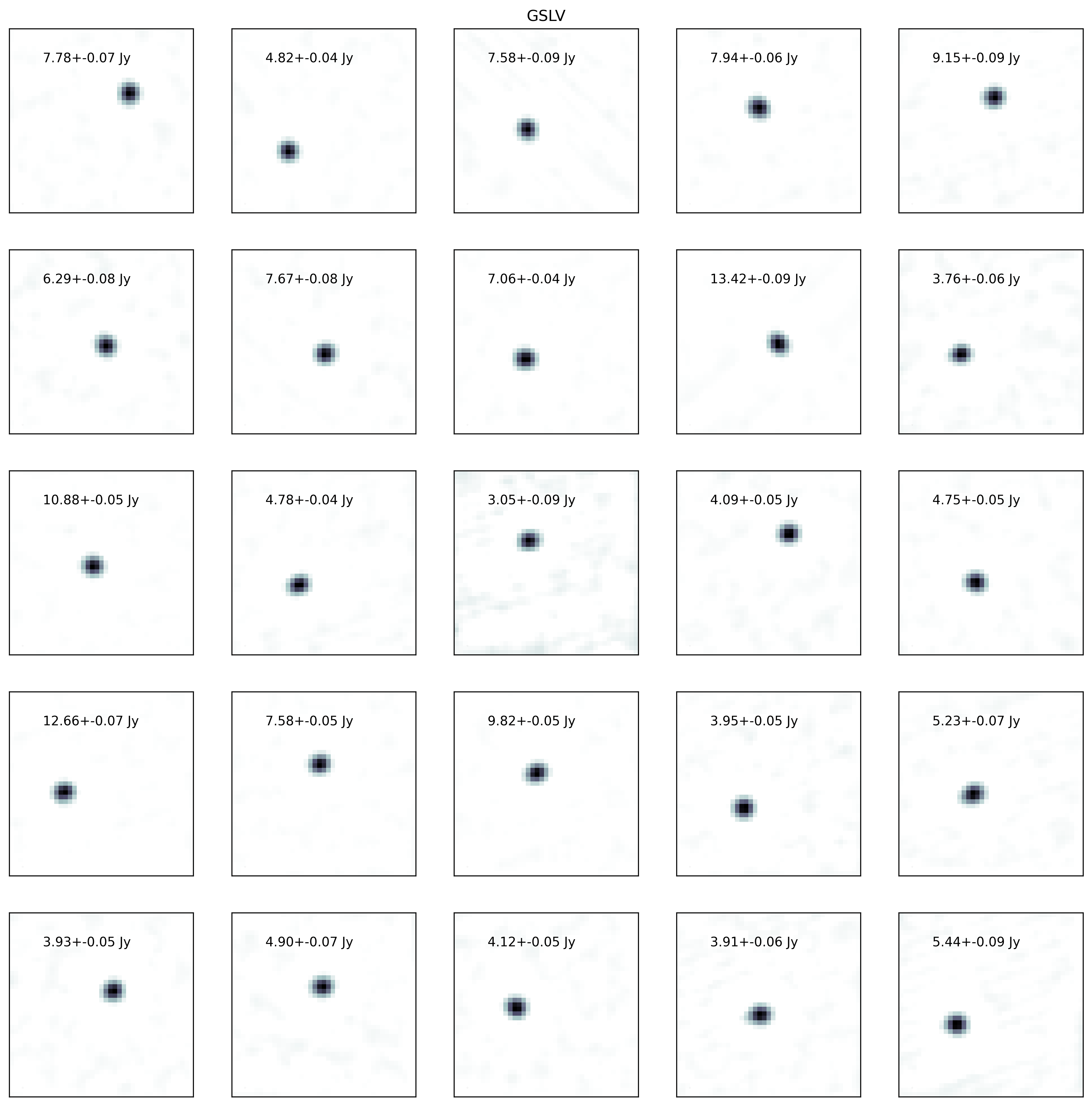}
    \caption{Same as Figure \ref{fig:starlink_gridmap} but for the two LVM3 rocket bodies discussed in Section \ref{sec:lvm3_thermal_emission}.}
    \label{fig:lvm3_gridmap}
\end{figure}

\end{document}

%% file: authors_oja.tex
\def\PhysicsPrinceton{1}
\def\UChicago{2}
\def\FNAL{3}
\def\KICPChicago{4}
\def\AAUChicago{5}
\def\Melbourne{6}
\def\IAP{7}
\def\ANLHEP{8}
\def\KIPAC{9}
\def\Stanford{10}
\def\SLAC{11}
\def\EFIChicago{12}
\def\PhysicsUChicago{13}
\def\Berkeley{14}
\def\ILAst{15}
\def\KEK{16}
\def\McGill{17}
\def\CIFAR{18}
\def\ColoradoAPS{19}
\def\ILPhys{20}
\def\UCLA{21}
\def\MSU{22}
\def\UCDavis{23}
\def\physicsNU{24}
\def\CASA{25}
\def\ColoradoPhys{26}
\def\CaseWestern{27}
\def\ILNCSA{28}
\def\Dunlap{29}
\def\UToronto{30}
\def\CfA{31}
\def\UNM{32}

\author{
  \href{http://orcid.org/0000-0002-7145-1824}{A.~Foster}\altaffilmark{\PhysicsPrinceton},
  A.~Chokshi\altaffilmark{\UChicago},
  \href{http://orcid.org/0000-0002-4435-4623}{A.~J.~Anderson}\altaffilmark{\FNAL,\KICPChicago,\AAUChicago},
  B.~Ansarinejad\altaffilmark{\Melbourne},
  \href{http://orcid.org/0000-0002-0517-9842}{M.~Archipley}\altaffilmark{\KICPChicago,\AAUChicago},
  \href{http://orcid.org/0000-0001-6899-1873}{L.~Balkenhol}\altaffilmark{\IAP},
  \href{http://orcid.org/0000-0002-1623-5651}{D.R.~Barron}\altaffilmark{\UNM},
  K.~Benabed\altaffilmark{\IAP},
  \href{http://orcid.org/0000-0001-5868-0748}{A.~N.~Bender}\altaffilmark{\ANLHEP,\KICPChicago,\AAUChicago},
  \href{http://orcid.org/0000-0002-5108-6823}{B.~A.~Benson}\altaffilmark{\FNAL,\KICPChicago,\AAUChicago},
  \href{http://orcid.org/0000-0003-4847-3483}{F.~Bianchini}\altaffilmark{\KIPAC,\Stanford,\SLAC},
  \href{http://orcid.org/0000-0001-7665-5079}{L.~E.~Bleem}\altaffilmark{\ANLHEP,\KICPChicago},
  \href{http://orcid.org/0000-0002-8051-2924}{F.~R.~Bouchet}\altaffilmark{\IAP},
  L.~Bryant\altaffilmark{\EFIChicago},
  \href{http://orcid.org/0000-0003-3483-8461}{E.~Camphuis}\altaffilmark{\IAP},
  J.~E.~Carlstrom\altaffilmark{\KICPChicago,\EFIChicago,\PhysicsUChicago,\ANLHEP,\AAUChicago},
  C.~L.~Chang\altaffilmark{\ANLHEP,\KICPChicago,\AAUChicago},
  P.~Chaubal\altaffilmark{\Melbourne},
  \href{http://orcid.org/0000-0002-5397-9035}{P.~M.~Chichura}\altaffilmark{\PhysicsUChicago,\KICPChicago},
  \href{http://orcid.org/0000-0002-3091-8790}{T.-L.~Chou}\altaffilmark{\AAUChicago,\KICPChicago},
  A.~Coerver\altaffilmark{\Berkeley},
  \href{http://orcid.org/0000-0001-9000-5013}{T.~M.~Crawford}\altaffilmark{\KICPChicago,\AAUChicago},
  \href{http://orcid.org/0000-0002-3760-2086}{C.~Daley}\altaffilmark{\ILAst},
  T.~de~Haan\altaffilmark{\KEK},
  K.~R.~Dibert\altaffilmark{\AAUChicago,\KICPChicago},
  M.~A.~Dobbs\altaffilmark{\McGill,\CIFAR},
  A.~Doussot\altaffilmark{\IAP},
  \href{http://orcid.org/0000-0002-9962-2058}{D.~Dutcher}\altaffilmark{\PhysicsPrinceton},
  W.~Everett\altaffilmark{\ColoradoAPS},
  C.~Feng\altaffilmark{\ILPhys},
  \href{http://orcid.org/0000-0002-4928-8813}{K.~R.~Ferguson}\altaffilmark{\UCLA,\MSU},
  K.~Fichman\altaffilmark{\PhysicsUChicago,\KICPChicago},
  S.~Galli\altaffilmark{\IAP},
  A.~E.~Gambrel\altaffilmark{\KICPChicago},
  R.~W.~Gardner\altaffilmark{\EFIChicago},
  F.~Ge\altaffilmark{\UCDavis},
  N.~Goeckner-Wald\altaffilmark{\Stanford,\KIPAC},
  \href{http://orcid.org/0000-0003-4245-2315}{R.~Gualtieri}\altaffilmark{\physicsNU},
  F.~Guidi\altaffilmark{\IAP},
  S.~Guns\altaffilmark{\Berkeley},
  N.~W.~Halverson\altaffilmark{\CASA,\ColoradoPhys},
  E.~Hivon\altaffilmark{\IAP},
  \href{http://orcid.org/0000-0002-0463-6394}{G.~P.~Holder}\altaffilmark{\ILPhys},
  W.~L.~Holzapfel\altaffilmark{\Berkeley},
  J.~C.~Hood\altaffilmark{\KICPChicago},
  A.~Hryciuk\altaffilmark{\PhysicsUChicago,\KICPChicago},
  N.~Huang\altaffilmark{\Berkeley},
  F.~K\'eruzor\'e\altaffilmark{\ANLHEP},
  A.~R.~Khalife\altaffilmark{\IAP},
  L.~Knox\altaffilmark{\UCDavis},
  M.~Korman\altaffilmark{\CaseWestern},
  K.~Kornoelje\altaffilmark{\AAUChicago,\KICPChicago},
  C.-L.~Kuo\altaffilmark{\KIPAC,\Stanford,\SLAC},
  K.~Levy\altaffilmark{\Melbourne},
  A.~E.~Lowitz\altaffilmark{\KICPChicago},
  C.~Lu\altaffilmark{\ILPhys},
  A.~Maniyar\altaffilmark{\KIPAC,\Stanford,\SLAC},
  E.~S.~Martsen\altaffilmark{\AAUChicago,\KICPChicago},
  F.~Menanteau\altaffilmark{\ILAst,\ILNCSA},
  \href{http://orcid.org/0000-0001-7317-0551}{M.~Millea}\altaffilmark{\Berkeley},
  J.~Montgomery\altaffilmark{\McGill},
  Y.~Nakato\altaffilmark{\Stanford},
  T.~Natoli\altaffilmark{\KICPChicago},
  \href{http://orcid.org/0000-0002-5254-243X}{G.~I.~Noble}\altaffilmark{\Dunlap,\UToronto},
  Y.~Omori\altaffilmark{\AAUChicago,\KICPChicago},
  \href{http://orcid.org/0000-0002-6164-9861}{Z.~Pan}\altaffilmark{\ANLHEP,\KICPChicago,\PhysicsUChicago},
  P.~Paschos\altaffilmark{\EFIChicago},
  \href{http://orcid.org/0000-0001-7946-557X}{K.~A.~Phadke}\altaffilmark{\ILAst,\ILNCSA},
  A.~W.~Pollak\altaffilmark{\UChicago},
  K.~Prabhu\altaffilmark{\UCDavis},
  W.~Quan\altaffilmark{\PhysicsUChicago,\KICPChicago},
  \href{http://orcid.org/0000-0003-1405-378X}M.~Rahimi\altaffilmark{\Melbourne},
  \href{http://orcid.org/0000-0003-3953-1776}{A.~Rahlin}\altaffilmark{\AAUChicago,\KICPChicago},
  \href{http://orcid.org/0000-0003-2226-9169}{C.~L.~Reichardt}\altaffilmark{\Melbourne},
  M.~Rouble\altaffilmark{\McGill},
  J.~E.~Ruhl\altaffilmark{\CaseWestern},
  E.~Schiappucci\altaffilmark{\Melbourne},
  \href{http://orcid.org/0000-0001-6155-5315}{J.~A.~Sobrin}\altaffilmark{\FNAL,\KICPChicago},
  A.~A.~Stark\altaffilmark{\CfA},
  J.~Stephen\altaffilmark{\EFIChicago},
  C.~Tandoi\altaffilmark{\ILAst},
  B.~Thorne\altaffilmark{\UCDavis},
  C.~Trendafilova\altaffilmark{\ILNCSA},
  \href{http://orcid.org/0000-0002-6805-6188}{C.~Umilta}\altaffilmark{\ILPhys},
  J.~D.~Vieira\altaffilmark{\ILAst,\ILPhys,\ILNCSA},
  A.~Vitrier\altaffilmark{\IAP},
  Y.~Wan\altaffilmark{\ILAst,\ILNCSA},
  \href{http://orcid.org/0000-0002-3157-0407}{N.~Whitehorn}\altaffilmark{\MSU},
  \href{http://orcid.org/0000-0001-5411-6920}{W.~L.~K.~Wu}\altaffilmark{\KIPAC,\SLAC},
  M.~R.~Young\altaffilmark{\FNAL,\KICPChicago},
  and
  J.~A.~Zebrowski\altaffilmark{\KICPChicago,\AAUChicago,\FNAL}
}

\altaffiltext{\PhysicsPrinceton}{Joseph Henry Laboratories of Physics, Jadwin Hall, Princeton University, Princeton, NJ 08544, USA}
\altaffiltext{\UChicago}{University of Chicago, 5640 South Ellis Avenue, Chicago, IL, 60637, USA}
\altaffiltext{\FNAL}{Fermi National Accelerator Laboratory, MS209, P.O. Box 500, Batavia, IL, 60510, USA}
\altaffiltext{\KICPChicago}{Kavli Institute for Cosmological Physics, University of Chicago, 5640 South Ellis Avenue, Chicago, IL, 60637, USA}
\altaffiltext{\AAUChicago}{Department of Astronomy and Astrophysics, University of Chicago, 5640 South Ellis Avenue, Chicago, IL, 60637, USA}
\altaffiltext{\Melbourne}{School of Physics, University of Melbourne, Parkville, VIC 3010, Australia}
\altaffiltext{\IAP}{Sorbonne Universit'e, CNRS, UMR 7095, Institut d'Astrophysique de Paris, 98 bis bd Arago, 75014 Paris, France}
\altaffiltext{\ANLHEP}{High-Energy Physics Division, Argonne National Laboratory, 9700 South Cass Avenue., Lemont, IL, 60439, USA}
\altaffiltext{\KIPAC}{Kavli Institute for Particle Astrophysics and Cosmology, Stanford University, 452 Lomita Mall, Stanford, CA, 94305, USA}
\altaffiltext{\Stanford}{Department of Physics, Stanford University, 382 Via Pueblo Mall, Stanford, CA, 94305, USA}
\altaffiltext{\SLAC}{SLAC National Accelerator Laboratory, 2575 Sand Hill Road, Menlo Park, CA, 94025, USA}
\altaffiltext{\EFIChicago}{Enrico Fermi Institute, University of Chicago, 5640 South Ellis Avenue, Chicago, IL, 60637, USA}
\altaffiltext{\PhysicsUChicago}{Department of Physics, University of Chicago, 5640 South Ellis Avenue, Chicago, IL, 60637, USA}
\altaffiltext{\Berkeley}{Department of Physics, University of California, Berkeley, CA, 94720, USA}
\altaffiltext{\ILAst}{Department of Astronomy, University of Illinois Urbana-Champaign, 1002 West Green Street, Urbana, IL, 61801, USA}
\altaffiltext{\KEK}{High Energy Accelerator Research Organization (KEK), Tsukuba, Ibaraki 305-0801, Japan}
\altaffiltext{\McGill}{Department of Physics and McGill Space Institute, McGill University, 3600 Rue University, Montreal, Quebec H3A 2T8, Canada}
\altaffiltext{\CIFAR}{Canadian Institute for Advanced Research, CIFAR Program in Gravity and the Extreme Universe, Toronto, ON, M5G 1Z8, Canada}
\altaffiltext{\ColoradoAPS}{Department of Astrophysical and Planetary Sciences, University of Colorado, Boulder, CO, 80309, USA}
\altaffiltext{\ILPhys}{Department of Physics, University of Illinois Urbana-Champaign, 1110 West Green Street, Urbana, IL, 61801, USA}
\altaffiltext{\UCLA}{Department of Physics and Astronomy, University of California, Los Angeles, CA, 90095, USA}
\altaffiltext{\MSU}{Department of Physics and Astronomy, Michigan State University, East Lansing, MI 48824, USA}
\altaffiltext{\UCDavis}{Department of Physics \& Astronomy, University of California, One Shields Avenue, Davis, CA 95616, USA}
\altaffiltext{\physicsNU}{Department of Physics and Astronomy, Northwestern University, 633 Clark St, Evanston, IL, 60208, USA}
\altaffiltext{\CASA}{CASA, Department of Astrophysical and Planetary Sciences, University of Colorado, Boulder, CO, 80309, USA }
\altaffiltext{\ColoradoPhys}{Department of Physics, University of Colorado, Boulder, CO, 80309, USA}
\altaffiltext{\CaseWestern}{Department of Physics, Case Western Reserve University, Cleveland, OH, 44106, USA}
\altaffiltext{\ILNCSA}{Center for AstroPhysical Surveys, National Center for Supercomputing Applications, Urbana, IL, 61801, USA}
\altaffiltext{\Dunlap}{Dunlap Institute for Astronomy \& Astrophysics, University of Toronto, 50 St. George Street, Toronto, ON, M5S 3H4, Canada}
\altaffiltext{\UToronto}{David A. Dunlap Department of Astronomy \& Astrophysics, University of Toronto, 50 St. George Street, Toronto, ON, M5S 3H4, Canada}
\altaffiltext{\CfA}{Harvard-Smithsonian Center for Astrophysics, 60 Garden Street, Cambridge, MA, 02138, USA}
\altaffiltext{\UNM}{Department of Physics and Astronomy, University of New Mexico, Albuquerque, NM, USA 87131}